\documentclass[useAMS,usenatbib]{mn2e}
\usepackage[dvips]{lscape,graphics,graphicx}
\usepackage{picture}
\usepackage{url}
\usepackage{amssymb,amsmath}
\usepackage{subfig}
\usepackage{color} 

\newcommand{\gtabouteq}{\,\hbox{\raise 0.5 ex \hbox{$>$}\kern-.77em 
                    \lower 0.5 ex \hbox{$\sim$}$\,$}}       
\newcommand{\ltabouteq}{\,\hbox{\raise 0.5 ex \hbox{$<$}\kern-.77em 
                     \lower 0.5 ex \hbox{$\sim$}$\,$}}

\title[The 617 MHz - $\lambda\,850~\mu$m Correlation in NGC~3044 and 
NGC~4157]{The 617 MHz - $\lambda\,850~\mu$m Correlation (Cosmic Rays and Cold Dust) in NGC~3044 and NGC~4157}
\author[J. A. Irwin, R. S. Brar, D. J. Saikia, \& R. N. Henriksen]{J. A. Irwin$^{1}$\thanks{E-mail:
irwin@astro.queensu.ca (JAI); Rupinder.Brar@uoit.ca (RSB); djs@ncra.tifr.res.in (DJS); henriksn@astro.queensu.ca (RNH)},
 R. S. Brar$^{2}$, D. J. Saikia$^{3}$ and R. N. Henriksen$^{1}$
\\
$^{1}$Dept. of Physics, Engineering Physics, \& Astronomy, Queen's University, Kingston, Ontario, Canada, K7L 3N6\\
$^{2}$Faculty of Science, University of Ontario Institute of Technology, 2000 Simcoe Street North, Oshawa, Ontario, Canada L1H 7K4\\
$^{3}$National Centre for Radio Astrophysics, TIFR, Pune University Campus, Post Bag 3, Pune, 411 007, India;
Cotton College State\\ University, Panbazar, Guwahati, 781 001, India
}
\begin{document}

\date{Accepted.... Received ...}

\pagerange{\pageref{firstpage}--\pageref{lastpage}} \pubyear{2002}

\maketitle

\label{firstpage}

\begin{abstract}
We present the first maps of NGC~3044 and NGC~4157 at $\lambda\,450~\mu$m and
 $\lambda\,850~\mu$m from the JCMT as well as
the first maps at 617 MHz from the GMRT.  High latitude
emission has been detected in both the radio continuum and sub-mm for
NGC~3044 and in the radio continuum for NGC~4157, including several new features.
For NGC~3044, in addition, we find 617 MHz emission extending to the north of the major
axis, beginning at the far ends of the major axis.  One of these low intensity features, more than
10 kpc from the major axis, has apparently associated emission at $\lambda\,20$ cm and
may be a result of
in-disk activity related to star formation. 

The dust spectrum at long wavelengths required fitting with a two-temperature model
for both galaxies, implying the presence of cold dust (T$_c\,=\,9.5$ K for NGC~3044 and
T$_c\,=\,15.3$ K for NGC~4157).  Dust masses are $M_d\,=\,1.6\,\times\,10^8~M_\odot$
and  $M_d\,=\,2.1\,\times\,10^7~M_\odot$ for NGC~3044 and NGC~4157, respectively,
and are dominated by the cold component.

There is a clear correlation between the 617 MHz and  $\lambda\,850~\mu$m emission in the
two galaxies.  In the case of NGC~3044 for which the  $\lambda\,850~\mu$m data are strongly
dominated by cold dust, this implies a relation between the non-thermal synchrotron emission
and cold dust.  The 617 MHz component represents an integration of massive star formation 
 over the past $10^{7-8}$ yr and the $\lambda\,850~\mu$m emission represents
 heating from the diffuse interstellar radiation
field (ISRF).

{The  617 MHz --  $\lambda\,850~\mu$m correlation improves when} a smoothing kernel is applied
to the  $\lambda\,850~\mu$m data to account for differences between the CR electron diffusion scale
and the mean free path of an ISRF photon to dust.  The best-fit relation is
$L_{617_{\rm MHz}}\,\propto\,{L_{850~\mu{\rm m}}}^{2.1\,\pm\,0.2}$ for NGC~3044.  
If variations in the cold dust emissivity are
dominated by variations in dust density, and the synchrotron emission depends on magnetic field strength 
(a function of gas density) as well as
CR electron generation (a function of massive star formation rate and therefore density via the Schmidt law)
then the expected correlation for NGC~3044 is $L_{617_{\rm MHz}}\,\propto\,{L_{850~\mu{\rm m}}}^{2.2}$, in agreement
with the observed correlation.

\end{abstract}


\section{Introduction}
\label{introduction}


\subsection{Cosmic Rays and Dust in Disk-Halo Outflows}
\label{sec:cr_dust}

Observations of edge-on, star-forming galaxies indicate that many
(possibly most) such systems display extra-planar emission (`halos').  Where
sufficient observations exist, these halos are known to be both multi-phase,
displaying all of the components of the in-disk interstellar medium (ISM),
as well as highly structured, revealing vertical features over a variety of spatial
scales
\citep[e.g.][]{lee01}. Mass flux estimates thus far 
\citep{bre94,wan95,wan01,fra02}
imply that disk-halo outflow is responsible for
transporting large quantities of gas and, as a result, has an important role
to play in the evolution of galaxies, from metallicity gradients in disks 
to altered star formation rates (SFRs). Spatially
resolved observations of nearby edge-on galaxies provide us with important data
for studying the details of such disk-halo outflows.

 Early observations of edge-on galaxies 
showed that radio continuum
emission can extend several kpc above the plane
\citep{hum89,hum91},
and this emission is
dominated by
non-thermal synchrotron radiation \citep[e.g.][]{irw99,irw00}, indicative of
magnetic fields and cosmic rays (CRs).  Galactic winds may also be present, since
CRs can drive outflows, as first pointed out by 
\cite{ipa75}; such outflows or `feedback' are crucial to galaxy formation
scenarios.

Dust in galaxy halos was first observed in the
form of vertical filaments in absorption against starlight \citep{sof87} and,
later in emission from
 high resolution space-based infra-red (IR) observations
\citep[e.g.][]{irw06}. Extracting the physical conditions of dust, which
is present over a range of sizes, compositions and excitation conditions, 
requires careful modelling
of the spectral energy distribution (SED) and therefore observations at
a variety of infrared wavelengths.  Although such modelling is routinely carried
out for galaxy disks, it has proved more difficult for the weaker halo
emission [but see \cite{wha09}]. With fewer data points, however, it is still possible
to constrain some dust properties \citep[e.g.][]{dun00,dun01}, especially in the `classical'
regime in which grains reach equilibrium temperatures.

In this paper, we present the results of observations 
at 617 MHz as well as at two sub-mm wavelengths,
$\lambda\,450$ $\mu$m and $\lambda\,850$ $\mu$m, of two edge-on galaxies --
NGC~3044 and NGC~4157 -- in order to study both the in-disk emission and any
halo emission which might be present.
Since synchrotron emission
is stronger at lower frequencies, halos should be more easily detected than
at high radio frequencies
 for comparable signal-to-noise (S/N).  A clear example
can be found in \cite{irw03}.
 As for dust, the sub-mm wavelengths are known to be probes of cold dust and
 are less likely to be contaminated by contributions from hot or warm components
and/or stochastically heated very small grains.  Such data also 
provide the best estimate of total dust mass. Adopting these wavebands also allows us to
probe the far infrared (FIR) - radio continuum relation (see next subsection) in these galaxies.

In the next two subsections, we discuss the FIR - radio continuum relation and introduce the
two galaxies.  In Sect.~\ref{sec:observations}, we provide details of the observations and data reduction.
Sect.~\ref{sec:results} presents the results, including high latitude emission, where present, 
temperature fitting, mass estimates, and sub-mm-radio correlations.  Sects.~\ref{sec:discussion} and
\ref{sec:conclusions} present the discussion and conclusions, respectively.

\subsection{The FIR-Radio Continuum Relation}
\label{sec:fir-rad}

The well-known relationship between thermal FIR
emission from dust and the predominantly non-thermal synchrotron emission
in galaxies \citep{dej85,hel85} is believed to result from the
dependence of both on
massive star formation
\citep{wun88}.  The same stars that provide ultraviolet (UV) photons to
heat the dust are also destined to become supernovae, thus producing
 non-thermal radio emission.  What is still a puzzle, however, is
how the relation can be so tight over many orders of magnitude
among galaxies that have a range of properties, for example, 
differing ISM masses, dust masses, metallicities, SFRs
and possible differences as to how 
optically thick or optically thin galaxies are to 
UV photons and CR electrons.  
Similar arguments apply to the correlation which has also been found {\it within}
galaxies.
As a result, other possibilities have
been proposed for the origin of the relation,
for example coupling between the magnetic field strength and the gas
density; since it is well known that gas and dust density vary
together, one would expect higher synchrotron emission wherever the
dust emission is higher \citep{hel93,hoe98,gro03}.

One observational difficulty has been the traditional choice of the
1.4 GHz radio frequency
 (which includes a small but non-negligible thermal contribution)
and FIR emission at $\lambda\,60$ and   $\lambda\,100$ $\mu$m (which includes
contributions from both warm and cold dust).  An early attempt at
separating the various spectral contributions within M~31 
\citep{hoe98}
suggested that a good
relation still exists between non-thermal radio emission and
{\it cold} dust, 
 the former driven by supernovae and the latter heated by
the general interstellar radiation field (ISRF) rather than
solely by hot young stars, i.e. the direct link with star-forming regions is
less clear.  
\cite{pie03} later found a slightly
non-linear relation between the radio continuum and cold dust 
($L_{1.4 GHz} \,\propto\, L_{cold~dust}^{1.13}$)
 and suggest that there may be a non-linear dependence of
radio emission on SFR.  See also \cite{tab10a} and \cite{tab10b} for
further discussion.

A better approach is to observe the radio emission at a frequency that 
is strongly dominated by the non-thermal component, so that assumptions
and corrections for the thermal contribution are not required.  This is especially
important when searching for correlations within a given galaxy since the
thermal contribution can vary strongly from place to place. 
We have taken this approach by 
choosing the low radio frequency of 617 MHz.
Similarly, by searching for correlations with
 $\lambda\,850$ $\mu$m emission, the ISRF-heated cold
dust component is more likely to be isolated.  Such an approach has been
taken by \cite{bra03} who find a good correlation between these two wavebands
within the galaxy, 
 NGC~5775. 

An important refinement is to consider whether a smoothed
version of the sub-mm emission improves upon the relation, should one be found.  For example, 
\cite{bic90} proposed that, since the mean free path of a UV photon to
dust absorption is much less than the diffusion length of a typical CR electron,
we would expect that the FIR-radio continuum relation (or, in our case, the
617 MHz - 850 $\mu$m relation) will be improved if the FIR emission is smoothed
spatially.  
Such an improvement has been observed by Murphy et al. (2006a, 2006b, 2008, 2009).
We will examine this refinement for NGC~3044
and NGC~4157 as well.

\begin{figure*}
\includegraphics*[width=14cm]{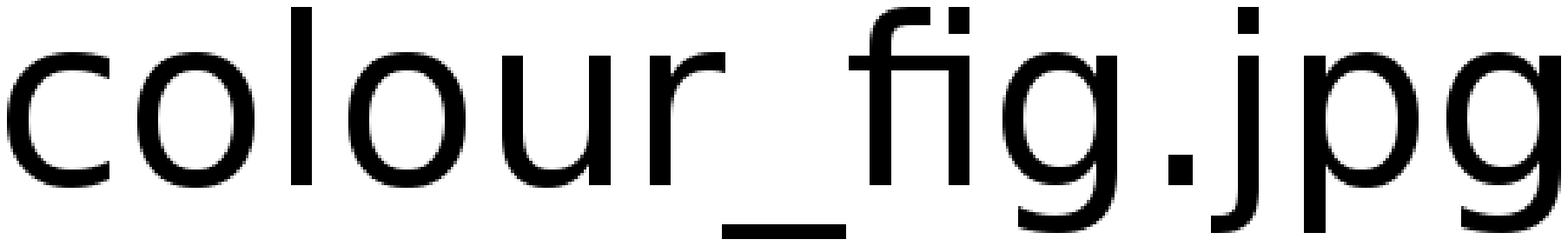}
  \caption{Colour images of NGC~3044 (left) and
NGC~4157 (right) from the Sloan Digital Sky Survey (SDSS).{\it This figure is included
as a jpeg file accompanying this manuscript.}}
\label{fig:colourfig}
\end{figure*}

\subsection{The galaxies}
\label{sec:thegalaxies}

Optical images of NGC~3044 and NGC~4157 are shown in Fig.~\ref{fig:colourfig} and their
basic 
 parameters 
are listed in Table~\ref{tab:galaxy_parameters}.  
Aside from the fact that NGC~4157 resides in a richer environment
(see below),
 the two galaxies have similar global properties: to within a factor of
about 1.5, the galaxies are of similar size, SFR, total mass, and gas
content.  One difference is in the colour of the emission as seen in
Fig.~\ref{fig:colourfig} and indicated by the (B-V)$_T$ colour of
Table~\ref{tab:galaxy_parameters}.
NGC~3044 shows predominantly blue emission
 throughout its projected disk, whereas NGC~4157
shows blue emission in the outer regions.  There is a small difference in inclination
between the two galaxies which may contribute to this colour difference though it isn't
clear to what extent.  For example, the slightly lower inclination of 
NGC~4157 appears to allow the redder inner regions of the galaxy to be more readily
revealed.
We describe these galaxies in detail, below.

\subsubsection{NGC~3044}

NGC~3044 (Fig.~\ref{fig:colourfig} left) 
is an isolated galaxy with no known companions.  
Although \cite{tul88} place it within the Leo Cloud, the other 3 members are widely displaced
in the sky ($>\,6^\circ$ or $>\,2.3$ Mpc). Some minor warping can be seen in the optical
image at large galactocentric radii,  however,
and asymmetries between the north-west and south-east major axis emission
have been noted by \cite{lee97} and \cite{col00}.
NGC~3044 has been considered both a `starburst' and `normal'
galaxy; \cite{tul06} 
discuss this issue, indicating
that it may be considered a starburst galaxy, but one in which the
starburst is extended over most of the disk, rather than concentrated 
near the nucleus. 

The presence of a galactic halo in NGC~3044 is well known, having
first been discovered in the radio continuum by \cite{hum89}
and further observed by various authors
\citep[e.g.][]{con90,col96,irw99,irw00}.
Extra-planar diffuse ionized gas (eDIG)
has  been detected by \cite{col00}, \cite{tul00} and \cite{mil03}
and extends as far as
5 kpc from the plane \citep{tul06}\footnote{Adjusted to our
distance and so throughout.}.  
Halo structure, including a kpc-scale diameter H$\alpha$ arc 
on the south side of the disk is also observed 
\citep{mil03}.
Ionization of the eDIG can be explained by
a combination of photo-ionization and shock ionization \citep{tul00}. 
XMM-Newton observations by \cite{tul06} 
have detected
an extended soft X-ray halo which reaches its maximum height over the
inner disk ($R\,<\,6$ kpc).  The X-ray emission correlates with both the 
eDIG and radio continuum.
Halo HI emission is also observed, including HI extensions and 
several loops
that have the signature of expanding
supershells \citep{lee97}. 
There appears to be some spatial correlation between the HI and radio
continuum features in the inner disk, but the HI is more extended 
radially and shows vertical extensions that do not have counterparts
at other wavebands.  Infrared spectroscopy of the halo has also been recently obtained by
\citet{ran11}.

A supernova was observed in NGC~3044 in the year, 1983\footnote{Data from
the International Astronomical Union Central Bureau for Astronomical Telegrams
at
http://www.cfa.harvard.edu/iau/lists/Supernovae.html.\label{footnote:iau}}.

\subsubsection{NGC~4157}

NGC~4157 (Fig.~\ref{fig:colourfig} right) is a member of the 
Ursa Major Cloud \citep{tul88} 
and belongs to the
galaxy group, LGG-258 \citep{gar93}. 
Previous radio continuum observations have been made by
\cite{con87}, \cite{gio87}, \cite{hum89}, \cite{irw99} and \cite{irw00}.
HI data can be found in \cite{rhe96}, in \cite{ver01}, 
in the Westerbork HI survey of Spiral and Irregular Galaxies
(WHISP) catalog\footnote{see
http://www.astro.rug.nl/\~\em  whisp/} 
\citep[see e.g.][]{noo05}, and in \cite{ken09}.
A companion galaxy, UGC~7176, with a dynamical mass $\approx\,$ 1\% of
that of NGC~4157, is located 12 arcmin to the south \citep{ken09}.
CO parameters can be found in 
\cite{you95} and \cite{kom08}.
The galaxy was observed with the Infrared Space Observatory (ISO)
at $\lambda\,12~\mu$m 
\citep{ben02} and
a UV image from the Galaxy Evolution Explorer
(GALEX) satellite can be found in 
\cite{gil07}.

A radio continuum halo was discovered in NGC~4157
by Irwin et al. (1999) but, 
to our knowledge, the only previous search for extraplanar
features in this galaxy was made by \cite{how99}
who searched for extraplanar dust absorption, with ambiguous results.

Three historical supernovae have been detected in NGC~4157
in the years, 1937, 1955 and 2003 (see Footnote~\ref{footnote:iau}).

\begin{table*}
 \centering
 \begin{minipage}{140mm}
  \caption{Galaxy Parameters}\label{tab:galaxy_parameters}
  \begin{tabular}{@{}lcc@{}}
  \hline
  Parameter\footnote{Values from the NASA/IPAC Extragalactic
Database (NED) unless otherwise indicated.}      & NGC~3044 & NGC~4157 \\
 \hline
Morphological Type & SB(s)c? sp  (HII)  & SAB(s)b? sp (HII)  \\
RA (J2000) (h m s) & 9 53 40.9 & 12 11 04.4\\
DEC (J2000) ($^\circ$ $^\prime$ $^{\prime\prime}$) &
1 34 47 & 50 29 05\\
Distance, $d$ (Mpc)\footnote{The quoted values agree, within errors, with
NED Hubble Flow distances with respect to the Cosmic Microwave
Background for $H_0$\,=\,73 km s$^{-1}$ Mpc$^{-1}$.}
& 21.7\footnote{\cite{lee97}.} & 
12.9\footnote{\cite{irw99}.} \\
Optical major $\times$ minor axis ($^\prime$ $\times$ $^\prime$) &
5.71 $\times$ 0.63 & 7.95 $\times$ 1.06\\
$D_{25}$ ($^\prime$, kpc)\footnote{Optical diameter at the 
25th magnitude isophote, from \cite{dev91}.} 
& 4.90, 30.9 & 6.76, 25.4 \\
Inclination ($^\circ$) & 85$^c$ & 83\footnote{Tully et al. (1996).}\\
Position angle ($^\circ$) & 113$^c$ & 66$^d$\\
Blue Magnitude & 12.46& 12.15\\
(B-V)$_T$\footnote{Total B-V colour from \cite{dev91}.}
 & $0.53\,\pm\,0.02$ & $0.80\,\pm\,0.01$ \\
$S_{12}$, $S_{25}$, $S_{60}$, $S_{100}$
(Jy)\footnote{IRAS 
flux densities at $\lambda\,12~\mu$m, $\lambda\,25~\mu$m, $\lambda\,60~\mu$m
and  $\lambda\,100~\mu$m, respectively, from \cite{san03}.}
  &  0.60, 1.11 9.64, 19.38 & 1.72, 2.12, 17.71, 50.67 \\
$\sigma_{12}$, $\sigma_{25}$, $\sigma_{60}$, $\sigma_{100}$
(mJy)\footnote{Errors on IRAS flux densities of previous row.}
  &  22, 86 30, 89 & 39, 28, 43, 170 \\
$S_{60}/S_{100}$ & 0.497& 0.350 \\
$F_{FIR}$, $F_{IR}$
($10^{-9}$ erg s$^{-1}$ cm$^{-2}$)\footnote{FIR ($\lambda\,40\,\to\,500~\mu$m)
 and IR ($\lambda\,8\,\to\,1000~\mu$m)
flux density, respectively, according
to the prescription of \cite{san96}, with their constant,
$C\,=\,1$.} & 0.558,
1.05 & 
 1.21, 2.35\\
$L_{FIR}$, $L_{IR}$ ($10^{10}$ $L_\odot$)\footnote{FIR and IR luminosity, 
respectively, from
$L\,=\,4\,\pi\,d^2\,F$, with $L_\odot\,=\,3.83\,\times\,10^{33}$ erg s$^{-1}$.} 
& 0.820, 1.55 & 
 0.631, 1.22\\
$L_{FIR}/D_{25}^2$, $L_{IR}/D_{25}^2$ 
($10^{40}$ erg s$^{-1}$ kpc$^{-2}$)
& 3.29, 6.21 & 3.76, 7.26 \\
SFR (M$_\odot$ yr$^{-1}$)\footnote{Star formation rate, 
from $L_{IR}$, according to the
prescription of \cite{ken98}.} & 2.7  & 2.1 \\
$M_{HI}$ ($10^{9}\,M_\odot$)\footnote{HI mass 
(Lee \& Irwin 1997\nocite{lee97} for NGC~3044; 
Kennedy 2009\nocite{ken09} for NGC~4157).}
& 5.4 $\pm$ 0.4 & 4.2 $\pm$ 0.1\\
$M_{H_2}$ ($10^{9}\,M_\odot$)\footnote{$H_2$ mass using
a conversion factor of $X\,=\,2.0\,\times\,10^{20}$ mol cm$^{-2}$ 
(K km s$^{-1}$)$^{-1}$ (Solomon \& Sage 1988\nocite{sol88} for NGC~3044; 
Young et al. 1996\nocite{you96} 
for NGC~4157).} 
& 2.1 $\pm$ 50\% & 1.8 $\pm$ 30\%\\
$M_{dyn}$ ($10^{11}\,M_\odot$)\footnote{Total dynamical mass from HI data
(Lee \& Irwin 1997\nocite{lee97}
for NGC~3044; Kennedy 2009\nocite{ken09} for NGC~4157).} 
& 1.5 $\pm$ 0.2 & 2.0 $\pm$ 0.5 \\
\hline
\end{tabular}
\end{minipage}
\end{table*}


\begin{table*}
 \centering
 \begin{minipage}{140mm}
  \caption{GMRT Observations\label{gmrt_observations}}
  \begin{tabular}{@{}lcc@{}}
  \hline
        & NGC~3044 & NGC~4157 \\
 \hline
Date of Observation & 26 July 2002 & 27 July 2002\\
On-source Observing Time (min.) & 404 & 280\\
Primary Flux Density Calibrator & 3C~286 & 3C~286\\
Phase Calibrator & J0943-083 & J1146+399\\
No. of Spectral Channels & 128 & 128\\
Channel Width (kHz) & 125 & 125\\
Central Frequency (MHz)\footnote{After editing}  & 617.375 & 619.937\\
Total Bandwidth (MHz)$^a$ & 9.875 & 9.375 \\
Primary Beam FWHM (arcmin) & 47.4 & 47.2 \\

\hline
\end{tabular}
\end{minipage}
\end{table*}

\section[]{Observations \& Data Reduction}
\label{sec:observations}

\subsection[]{Radio Continuum Data}

Observations of NGC~3044 and NGC~4157 were obtained with the
Giant Metrewave Radio Telescope (GMRT), located near Pune, India,
which consists of 30 antennas, each 45 m in diameter, in a 
fixed, `Y'-shaped
array with a longest baseline of about 25 km and the shortest, 100 m.  
Fourteen of the antennas
are clustered randomly in a central 1 $\times$ 1 km central `square'
with the remainder in the arms.
At the time of the observations, not all antennas were available, leaving
typically, 25 to 29 antennas for each observation.
For a more complete description of the telescope, see
\cite{swa91a}, \cite{swa91b}, \cite{ana05} or
http://www.gmrt.ncra.tifr.res.in.
A summary of the observational data is given in Table~\ref{gmrt_observations}.
Note that the shortest spacing ensures that, at 617 MHz, spatial scales 
up to about 17 arcmin are detected, i.e. about 3 times and 2 times the
optical major axis diameters of NGC~3044 and NGC~4157, respectively.  

Observations were carried out in the standard fashion, with a flux
density calibrator observed during the 
observing run and phase calibrators observed
at regular intervals, typically every 30 minutes.  
The spectral line mode, which is the default at the GMRT, was used
to identify interference,
if present.  The data were processed using the Astronomical
Image Processing System (AIPS) of the National Radio Astronomy
Observatory (NRAO).  
Editing was initially carried out using the AIPS-compatible routine,
GMRED\footnote{This AIPS-compatible routine was written by us and is available at
{\tt http://www.astro.queensu.ca/$\,\tilde{\,}$irwin}.} which we
wrote in order to remove bad data, since the GMRT did not 
have
on-line flagging during the observations.  Further editing was
then carried out in AIPS
using standard routines. 
After editing, the central frequencies were 
slightly different
for the two galaxies (Table~\ref{gmrt_observations}),
but we will refer to both data sets as the 617 MHz data.

Standard continuum calibration was carried out except for the additional
step of
 bandpass calibration using the phase calibrators.
The uv data were then Fourier-transformed, the dirty beam deconvolved
and the clean beam reconvolved
using the AIPS routine, IMAGR.  This required forming a number of
fields so that significant sources far from the field center could
be properly imaged and cleaned.  Maps were originally 
formed for every channel
so that each one could be checked for consistency.  This led to
some further editing and re-calibration.  The uv data in the remaining
 channels
were then averaged to form a single-channel data set.  From these data,
maps were formed using a range of uv weightings resulting in
a range of spatial resolutions and signal-to-noise (S/N) ratios,
from which we present the 9 arcsec and 15 arcsec resolution images 
in Figs.~\ref{n3044resultsfig}a and b and
Figs.~\ref{n4157resultsfig}a and b. 
The undisplayed images do not reveal any features that are not seen
on these two.

The final step was to correct
for the primary beam of the GMRT.  This correction, however,
introduces
 position-dependent noise over the image and, since the 
total flux correction was measured to be less than 0.4\% for both galaxies
after the primary beam correction
(much less than other errors), we adopt the uncorrected maps for 
presentation and further analysis.

We carried out observations of NGC~3044 and NGC~4157 at 327 MHz also.
However, the resulting data sets were not of sufficient quality
to use.  Since we have previously obtained Very Large Array (VLA)
data of NGC~3044 and NGC~4157 at $\lambda\,20$ cm, 
these data are also presented here (Figs.~\ref{n3044vlafig} and \ref{n4157vlafig}, respectively) 
to supplement our GMRT data.  For details on these
data sets, see \cite{irw99}.

\subsection[]{Sub-mm Images}

Sub-mm observations of NGC~3044 and NGC~4157 were obtained with the
Submillimetre Common-User Bolometer Array (SCUBA, see 
\cite{hol99}) 
on the James
Clerk Maxwell Telescope (JCMT\footnote{The JCMT is supported by the
Science and Technology Facilities Council
of the UK, the NRC of Canada, and the
Netherlands Organization for Scientific Research.}).  
The $\lambda\,450~\mu$m and
 $\lambda\,850~\mu$m arrays, which were used concurrently,
consist of 91 and 37 circular pixels, respectively.  The configuration
of the pixels gives an instantaneous field of view of 2.3$^\prime$.
Observations were carried out in 64-point jiggle-map mode which ensures
Nyquist sampling at both frequencies.  Hourly pointing checks were
made during the observations and measurements to determine the
atmospheric transparency were also carried out.  Two fields of
view were required to cover each galaxy.  The final field of view
and other observing data are given in Table~\ref{jcmt_observations}.
We will refer to these data as the $\lambda\,450~\mu$m and
 $\lambda\,850~\mu$m data.

The data were reduced using the SCUBA User 
Reduction Facility (SURF)
package \citep{jen99}.
  The steps included
regridding to true sky positions, atmospheric subtraction,
flatfielding, and correction for sky opacity.
For the latter, the optical depth from
the Caltech Sub-mm Observatory, $\tau_{CSO}$, was used, extrapolated
to the observing wavelengths, and interpolated in time.  
 Since there are not many pixels with no galaxy emission,
sky values were calculated by first removing source emission from
the data prior to sky subtraction\footnote{The SURF routine, CALCSKY,
was used.}.  The data
were also examined carefully and edited for noisy bolometers and 
data spikes.
Corrections for positional drift were applied and the data were
calibrated using values of the known flux density calibrators
for the times of observation.  At this point, 
the data consisted of a sequence of calibrated images in time.
Each image was then
re-checked for bad data and additional flagging was carried
out if the pixel value deviated by more than $5\,\sigma$ from the mean
 pixel value from all images.  The final image was then made
from an average of all images, weighted by integration time
and background noise.
The final step involved blanking the periphery of the final field of
view (which contains two 2.3$^\prime$ fields) where the noise was higher
than the average over the general field.

Since these data were obtained, \cite{dif08} have independently reduced the
same data set with good agreement in the results. We will compare our results to theirs,
where relevant, in subsequent sections.

Finally, we consider the possibility of contamination of the $\lambda\,850$
$\mu$m band by CO(J=3-2) emission.  For NGC~3044, this can be tested explicitly
since we have CO(J=3-2) data from \citet{lee98} who find a maximum brightness
temperature of 0.091 K with an effective line width of 89 km s$^{-1}$ 
(width of $\Delta\,\lambda$ = 7.5 $\mu$m)
within a 14.9 arcsec beam
at a location
near the radio continuum peak.  Our peak continuum flux density of
69.4 mJy beam$^{-1}$ in a 15 arcsec beam (Fig.~\ref{n3044resultsfig}d) corresponds
to 2.1 K within the SCUBA 850 $\mu$m bandwidth of 70 $\mu$m \citep{hol99}.
Therefore, the estimated contamination is less than 1\%.  It is well known that
contamination of the SCUBA 450 $\mu$m band by CO(J=6-5) is even lower
\citep[e.g.][]{sea04}.

\begin{table*}
 \centering
 \begin{minipage}{140mm}
  \caption{JCMT Observations\label{jcmt_observations}}
  \begin{tabular}{@{}lcc@{}}
  \hline
        & NGC~3044 & NGC~4157 \\
 \hline
Dates of Observation & 5, 6 Jan. 2001 & 12, 13 Jan. 2003\\
                     & 15 Nov. 2001   & 11 Oct. 2003 \\
On-source Observing Time (min.) & 450 & 480\\
Flux Density Calibrators & Mars, CRL~618 & Uranus, Mars\\
Central Wavelength ($\mu$m) & 443, 863 & 443, 863 \\
Resolution ($^{\prime\prime}$)\footnote{At $\lambda\,450~\mu$m and
 $\lambda\,850~\mu$m, respectively.}
& 8.0, 14.9 & 8.0, 14.4 \\
Field of view (arcminute$^2$)$^a\,$\footnote{After editing.} & 
6.13, 8.0 &  6.48, 8.42 \\
\hline
\end{tabular}
\end{minipage}
\end{table*}

\section[]{Results}
\label{sec:results}

\begin{figure*}
\includegraphics*[width=15cm]{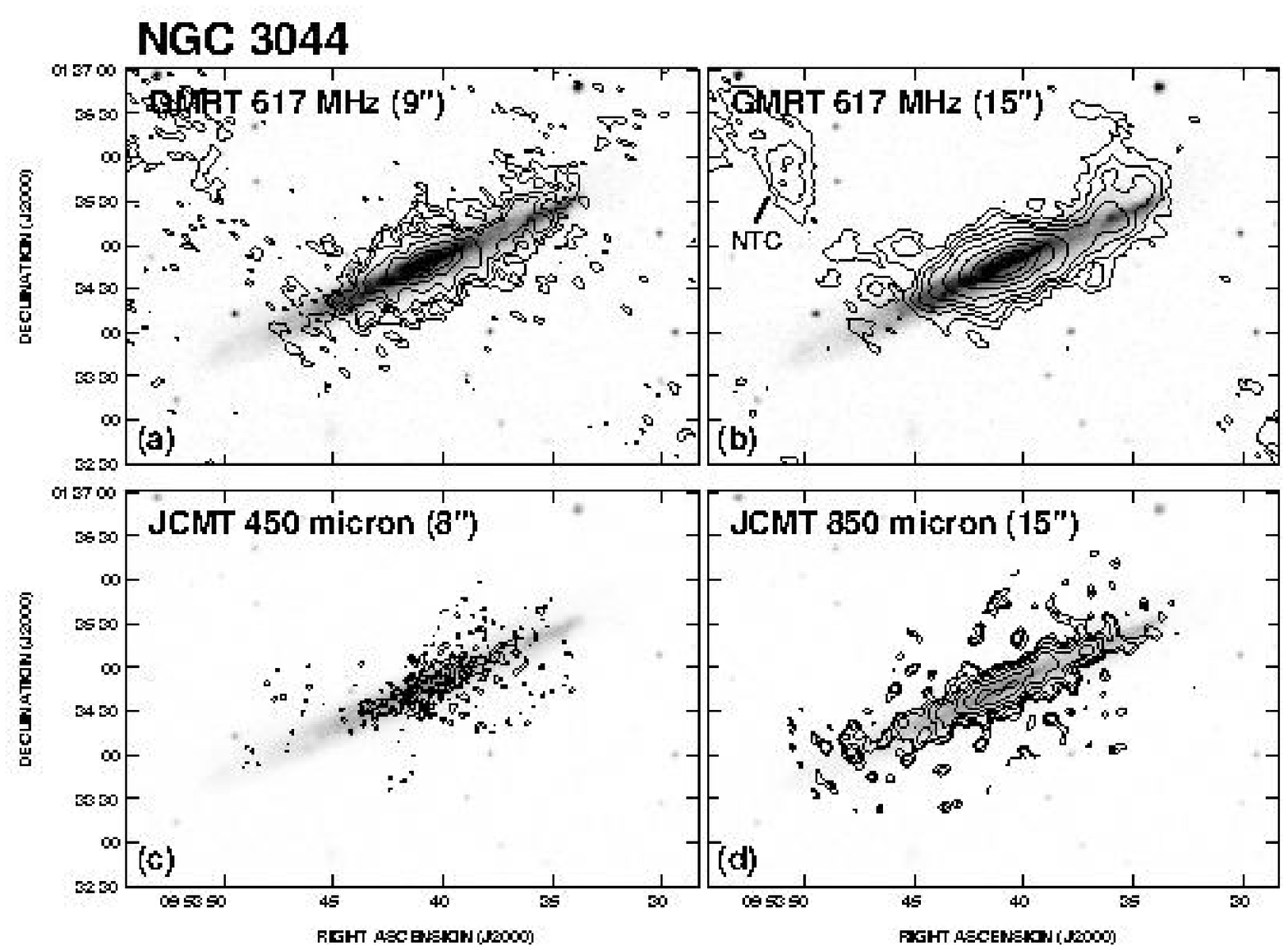}
  \caption{Radio continuum and sub-mm emission of NGC~3044 (contours)
superimposed on a greyscale of the Digitized Sky Survey 2 (DSS2)
blue image, whose grey range has been varied, as needed, to show the
contours.  In each case, the first contour has been set to $2\,\sigma$,
where $\sigma$ is the rms noise of the image. 
{\bf (a)} GMRT 617 MHz map with a resolution (circular beam FWHM 
and so throughout) of 
9.0$^{\prime\prime}$.  Contours are at 0.29, 0.5, 0.75, 1.5, 3,
5, 7.5, and 10 mJy beam$^{-1}$ and the map peak is
25.4 mJy beam$^{-1}$.  {\bf (b)}
GMRT 617 MHz map (resolution = 
15$^{\prime\prime}$).  Contours are at 0.24, 0.5, 0.9, 1.5, 3,
 6, 10, 15, and 21 mJy beam$^{-1}$ and the map peak is
27.4 mJy beam$^{-1}$. The non-thermal cloud (NTC) discussed in Sect.~\ref{sec:n3044-discuss}
is marked.
{\bf (c)}  JCMT $\lambda$ 450 $\mu$m map (resolution
= 8.0$^{\prime\prime}$).  Contours are at 64, 100, 150, 200, and 300
mJy beam$^{-1}$ and the map peak is 307.8 mJy beam$^{-1}$. {\bf (d)}
JCMT $\lambda$ 850 $\mu$m map (resolution
= 14.9$^{\prime\prime}$).  Contours are at 9, 13, 18, 25, 35,\
50 and 65 
mJy beam$^{-1}$ and the map peak is 69.4 mJy beam$^{-1}$.}
\label{n3044resultsfig}
\end{figure*}

The final maps
are shown in Fig.~\ref{n3044resultsfig} for NGC~3044.
and Fig.~\ref{n4157resultsfig} for NGC~4157.  
Beam sizes and noise value information can be found in the figure captions.
We discuss the results for
the two galaxies, separately.

\begin{figure*}
\includegraphics[width=15cm]{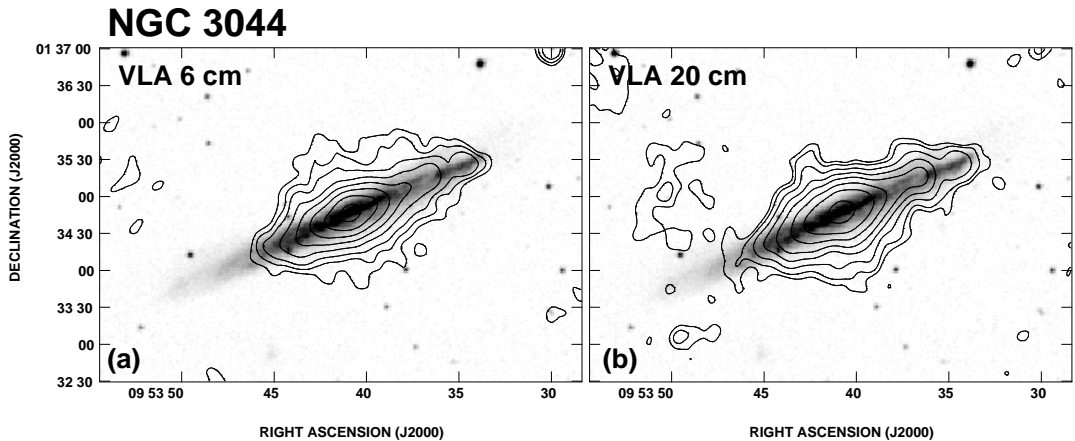}
  \caption{VLA radio continuum emission (contours) superimposed on
the DSS2 blue image (greyscale).  For more information, see
Irwin et al. (1999).  {\bf (a)} $\lambda\,6$ cm emission.  The beam is
18.8$^{\prime\prime}$ $\times$ 15.8$^{\prime\prime}$ at a position
angle of $\theta\,=\,$7.6$^\circ$ and contours are at
0.08 (2$\,\sigma$),
0.15, 0.25, 0.5, 1, 2, 4, 8, and 10 mJy beam$^{-1}$.  The map
peak is 10.2 mJy beam$^{-1}$.  {\bf (b)}
$\lambda\,20$ cm emission.  The beam is
18.9$^{\prime\prime}$ $\times$ 15.9$^{\prime\prime}$ at 
$\theta\,=\,$-7.1$^\circ$ and contours are at
0.12 (2$\,\sigma$), 
0.2, 0.4, 0.8, 1.5, 3, 6, 15, and 21 mJy beam$^{-1}$.  The map
peak is 21.7 mJy beam$^{-1}$.
}
\label{n3044vlafig}
\end{figure*}

\subsection{The NGC~3044 Maps}

\subsubsection{Radio Continuum Images}
\label{sec:radcont_images}

Figs.~\ref{n3044resultsfig}a and b show
617 MHz data for NGC~3044, illustrating both the
 in-disk as well as high latitude
radio continuum emission of this galaxy, and the flux density
is given in Table~\ref{properties_table}.
  The data have been shown with two 
different uv weightings to illustrate the fine structure emission 
(Fig.~\ref{n3044resultsfig}a) as well as larger scale structure
at higher sensitivity (Fig.~\ref{n3044resultsfig}b).  At 617 MHz,
the emission is strongly dominated, unambiguously,
by non-thermal synchrotron radiation.  For comparison, we show the
$\lambda\,6$ cm and $\lambda\,20$ cm VLA maps 
from \cite{irw99} in Fig.~\ref{n3044vlafig}
 over the same field of view as 
Fig.~\ref{n3044resultsfig}. {As shown by
 \cite{irw99}, the $\lambda\,6$ cm and $\lambda\,20$ cm maps are also dominated
by non-thermal emission.  In addition, we compute the global spectral index
between the  617 MHz and $\lambda\,20$ cm emission to be
$\alpha\,=\,-0.72\,\pm\,0.25$; this spectral index strongly departs from
what would be expected for thermal emission ($\alpha_{thermal}\,=\,-0.1$).}

The 617 MHz radio continuum emission in the disk of NGC~3044 is highly 
asymmetric,
being extended on the north-west side but has a more `truncated'
appearance on the south-east side.  Asymmetries, in just this sense, can be
seen in the X-ray image \citep{tul06} 
and in H$\,\alpha$ \citep[e.g.][]{col00}
as well as the $\lambda\,$6 cm and 
$\lambda\,$20 cm maps.
 As has been pointed out earlier 
(Sect.~\ref{introduction}), the X-ray, radio continuum, and H$\,\alpha$
all appear correlated.  The asymmetry observed in HI, however, does
not correlate in the same sense.  The HI distribution is extended more
on the south-east disk than in the north-west 
\citep{lee97}. 

The extraplanar radio continuum emission is highly structured, as has
been seen before in radio halos of other edge-on galaxies
\citep[e.g.][]{lee01}
 though the halo is not as pronounced
as in the VLA maps likely because the
VLA observations have somewhat higher dynamic range (255/1 and
362/1 for $\lambda\,$6 cm and $\lambda\,$20 cm, respectively) than
the 617 MHz observations (228/1).  The uv distribution will also differ.
The broadest halo can be seen on
either side of the nuclear region, consistent with what is seen in
H$\,\alpha$ and soft X-rays.  

A new and interesting result is the appearance of the
two far ends of the major axis, both of which show emission extending out
of the plane towards the north and both of which show a `double-pronged'
appearance.  That is, there is evidence for the ejection of cosmic rays
away from the plane at both ends of the major axis; presumably, the
ejection of particles is facilitated by the lower density ISM at these
locations.  


On the far south-east end of the major axis, moreover, where the radio continuum major axis
emission is truncated, some 
disconnected emission features can be seen towards the north, including a large feature centered
at 
RA $\approx$ 9$^{\rm h}$ 53$^{\rm m}$ 51$^{\rm s}$,
DEC $\approx$ 01$^\circ$ 35$^\prime$ 45$^{\prime\prime}$
which we have labelled a `non-thermal cloud' (NTC).  The feature occurs at a low
S/N, but the independent VLA 
$\lambda\,$20 cm map
and (to a lesser extent) in the VLA $\lambda\,$6 cm map
(Fig.~\ref{n3044vlafig}) also show features in this direction.  
These features will be discussed further in Sect.~\ref{sec:n3044-discuss}.

\subsubsection{Sub-mm Images}

The sub-mm emission of NGC~3044 is shown in Fig.~\ref{n3044resultsfig}c and d.
The field of view, delineated by the region over which contour
emission is displayed, is much smaller than the radio continuum emission 
(see Table~\ref{jcmt_observations} for the field size).
These maps show mainly the strong, in-disk emission, since the sensitivity
is insufficient 
 to delineate a global dust halo, if it exists.  The
$\lambda\,$850 $\mu$m map, however, does hint at a few discrete features;
these will be discussed in Sect.~\ref{sec:n3044-discuss}.

The total $\lambda\,$450 $\mu$m and  $\lambda\,$850 $\mu$m flux densities 
are given in Table~\ref{properties_table}. 
If emission at the level of the rms noise 
were to exist out to the ends of the optical major axis 
 in the $\lambda$450 $\mu$m map (as is the case for the $\lambda\,$850 $\mu$m map), then 
the $\lambda$450 $\mu$m flux density would not change by more
than the error bar that has been quoted. Therefore, the flux densities at the two frequencies
can both be used to constrain dust properties, within the constraints of their uncertainties,
as will be discussed in
Sect.~\ref{dust_properties}.

\begin{table}
 \centering
 \begin{minipage}{300mm}
  \caption{Measured and Derived Galaxy Properties\label{properties_table}}
  \begin{tabular}{@{}lcc@{}}
  \hline
 Property    & NGC~3044 & NGC~4157 \\
 \hline
$S_{617~\rm MHz}$ (Jy)\footnote{Flux density at 617 MHz.  
Uncertaintites include variations\\
 in 
editing, and choice of imaging and self-calibration parameters\\
 over different
reduction trials.}
 & 0.20 $\pm$ 0.04 & 0.43 $\pm$ 0.05\\
$S_{450~\mu \rm m}$ (Jy)\footnote{Flux density at 450 or 850 $\mu$m.
Uncertaintites include\\ 
variations over different box sizes as
well as variations between\\
 our data reduction
 and that of  \cite{dif08}.}
 & 3 $\pm$ 1 & 10 $\pm$ 2 \\
$S_{850~\mu \rm m}$ (Jy)$^b$ & 0.75 $\pm$ 0.10 & 0.8 $\pm$ 0.1\\
T$_{c}$ (K)\footnote{Temperature of cold (subscript, c, and so throughout)
or warm\\
 (subscript, w, and so throughout) dust component (see text).}
& 9.5 $\pm$ 1.5 & 15.3 $\pm$ 3 \\
T$_{w}$$^c$ (K) & 31.0 $\pm$ 1.4 & 25.6 $\pm$ 0.9 \\
N$_c$/N$_w$\footnote{Eqn.~\ref{two_fit_eqn}. The estimated error is $\approx$ 25\%.
This quantity is\\
 equivalent to $M_c/M_w$.} & 76 & 4 \\
$M_d$ ($10^8$ $M_\odot$)\footnote{Eqn.~\ref{dust_mass_eqn}.} 
&  1.6 $\pm$ 0.6 & 0.21 $\pm$ 0.06 \\
$M_{gas}/M_{dust}$\footnote{From $M_{HI}\,+\,M_{H_2}$ of 
Table~\ref{tab:galaxy_parameters} and $M_d$ from this table.  The\\
 uncertainty could be as high as
90\% for NGC~3044 and 60\%\\
 for NGC~4157.} & 47 & 286\\
\hline
\end{tabular}
\end{minipage}
\end{table}

\begin{figure*}
\includegraphics*[width=16cm]{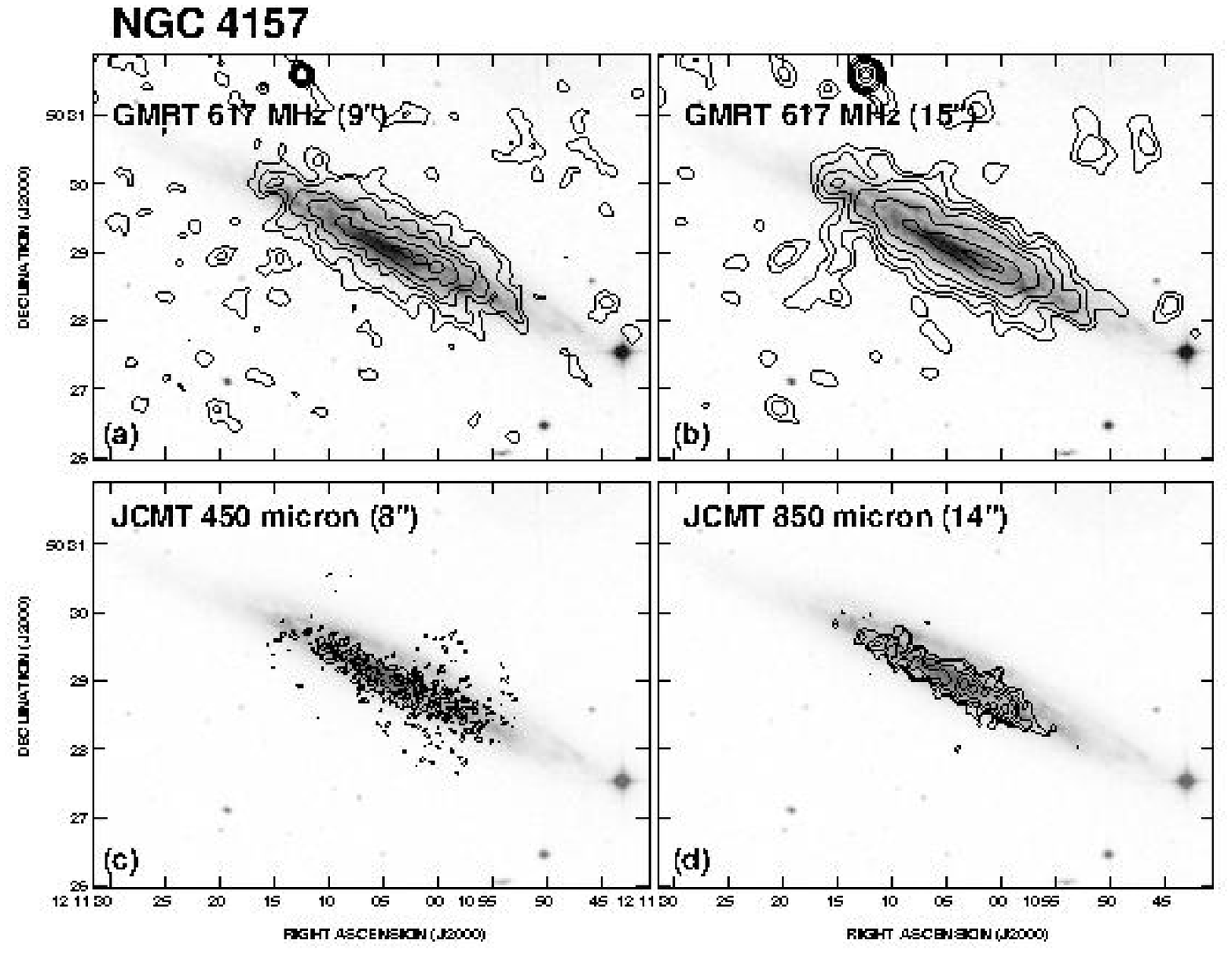}
  \caption{Radio continuum and sub-mm emission of NGC~4157 (contours)
superimposed on a greyscale of the Digitized Sky Survey 2 (DSS2)
blue image, whose grey range has been varied, as needed, to show the
contours.  In each case, the first contour has been set to $2\,\sigma$,
where $\sigma$ is the rms noise of the image. 
{\bf (a)} GMRT 617 MHz map with a resolution (circular beam FWHM 
and so throughout) of 
9.0$^{\prime\prime}$.  Contours are at 0.68, 1.2, 2, 3, 5,
 and 7 mJy beam$^{-1}$ and the map peak is
19.9 mJy beam$^{-1}$.  {\bf (b)}
GMRT 617 MHz map (resolution = 
15$^{\prime\prime}$).  Contours are at 0.88, 1.4, 2.3, 3.5, 6,
12, and 17 mJy beam$^{-1}$ and the map peak is
20.0 mJy beam$^{-1}$. {\bf (c)}  JCMT $\lambda$ 450 $\mu$m map (resolution
= 8.0$^{\prime\prime}$).  Contours are at 102, 170, 250, and 350
mJy beam$^{-1}$ and the map peak is 391 mJy beam$^{-1}$. {\bf (d)}
JCMT $\lambda$ 850 $\mu$m map (resolution
= 14.4$^{\prime\prime}$).  Contours are at 17, 25, 35, 45, and
and 55
mJy beam$^{-1}$ and the map peak is 62.2 mJy beam$^{-1}$.}
\label{n4157resultsfig}
\end{figure*}

\begin{figure}
\includegraphics[width=6.5cm]{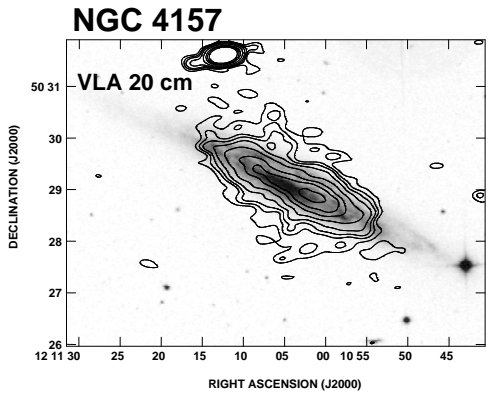}
  \caption{VLA radio continuum $\lambda\,$20 cm
emission (contours) superimposed on
the DSS2 blue image (greyscale).  For more information, see
Irwin et al. (1999).  The beam is
21.1$^{\prime\prime}$ $\times$ 11.3$^{\prime\prime}$ at 
$\theta\,=\,$-81.4$^\circ$ and contours are at
0.22 (2$\,\sigma$), 
0.4, 0.6, 1, 2, 4, 6 and 9 mJy beam$^{-1}$.  The map
peak is 44.1 mJy beam$^{-1}$.
}
\label{n4157vlafig}
\end{figure}

\subsection{The NGC~4157 Maps}

\subsubsection{Radio Continuum Images}
\label{sec:n4157_radio}

The 617 MHz emission of NGC~4157 is shown in Fig.~\ref{n4157resultsfig}a
and b and the flux density is given in Table~\ref{properties_table}.
 For comparison, the VLA $\lambda\,20$ cm image is shown in
Fig.~\ref{n4157vlafig} (a $\lambda\,$6 cm image does not yet exist). 
 The strongest radio emission belongs to a background
point source that can be seen to the north of the galaxy on the far east side.
The 617 MHz data agree very well with the VLA data throughout the disk.
Fig.~\ref{n4157resultsfig}a and b appear to show a slight offset between the
optical and radio major axes.  However, this is because of a north-west/south-east
assymetry on either side of the major axis of the DSS2
blue image that has been displayed in the figure; 
the south-east side is the
farther side and therefore the displayed optical blue emission is truncated by extinction on this side. 
The colour image of Fig.~\ref{fig:colourfig}, for example, shows good
north-west to south-east symmetry and the optical and radio major axes align well
when comparing with the colour image.

 The radio emission of the halo is also similar
between the VLA $\lambda\,20$ and 617 MHz emission, although the 
discrete extensions
seen at $\lambda\,20$ cm are less obvious at 617 MHz, possibly because 
of the
lower dynamic range at 617 MHz (42/1 over the emission associated
with NGC~4157 in Fig.~\ref{n4157resultsfig}b, compared to
93/1 for Fig.~\ref{n4157vlafig}).  The most obvious extension in
Fig.~\ref{n4157resultsfig}b begins
in the far north-eastern disk and extends approximately 1.5 arcmin (5.6 kpc) towards
the south.  It is visible at 617 MHz but not at $\lambda\,20$ cm.  
If this feature is real, {its spectral index must be less than}
 $\alpha\,=\,-1.7$ ($I_\nu\,\propto\,\nu^\alpha$) to render it
undetectable on the  $\lambda\,20$ cm image.  No other independent detection
of this feature has been made, but we note that it falls along the eastern
side of an HI extension detected by \citet{ken09}; also {spectral indices as low as}
 $\alpha\,\approx\,-2$ have previously been detected in vertical features related to
outflow in other galaxies \citep{hee11,irw12}.  See Sect.~\ref{sec:n4157-discuss} for further discussion.

\subsubsection{Sub-mm Images}
\label{sec:n4157_submm}

The $\lambda\,450~\mu$m and $\lambda\,850~\mu$m maps are shown in
Figs.~\ref{n4157resultsfig}c and d, respectively, and flux densities
are given in Table~\ref{properties_table}.
Only in-disk emission
can be seen in these full-resolution images.  
Again, there is the appearance of an offset in the major axis
between the sub-mm emission and the optical emission.  However,
brightest optical emission aligns well with the brightest sub-mm
emission, as shown in Figs.~\ref{n4157resultsfig}c and d to within
the estimated pointing accuracy of 3 arcsec\footnote{We have also verified that
the astrometry of the $\lambda\,450~\mu$m and $\lambda\,850~\mu$m maps
agrees with the results of \cite{dif08}
who have independently reduced the
SCUBA data.}, and, when
 the sub-mm emission maps are overlaid on the SDSS R-band image
(not shown), the emission again centers squarely on the brightest optical
emission. As we saw with the radio continuum emission, dust obscuration
creates an apparent asymmetry in the optical images 
that have been displayed in the overlays.

The $\lambda\,450~\mu$m and $\lambda\,850~\mu$m emission in NGC~4157 extends approximately
equally as far in radius; however, in both cases, neither is seen as far out as the optical
disk.  This is likely a sensitivity issue since the sub-mm dynamic range of the NGC~4157
data is lower than that of NGC~3044.  This must be kept in mind when the two galaxies
are compared. 


\subsection{Dust Temperature, Mass and Gas/Dust Ratio}
\label{dust_properties}

\begin{figure*}
\begin{minipage}{0.45\linewidth}
\centering
\includegraphics[width=7.5cm]{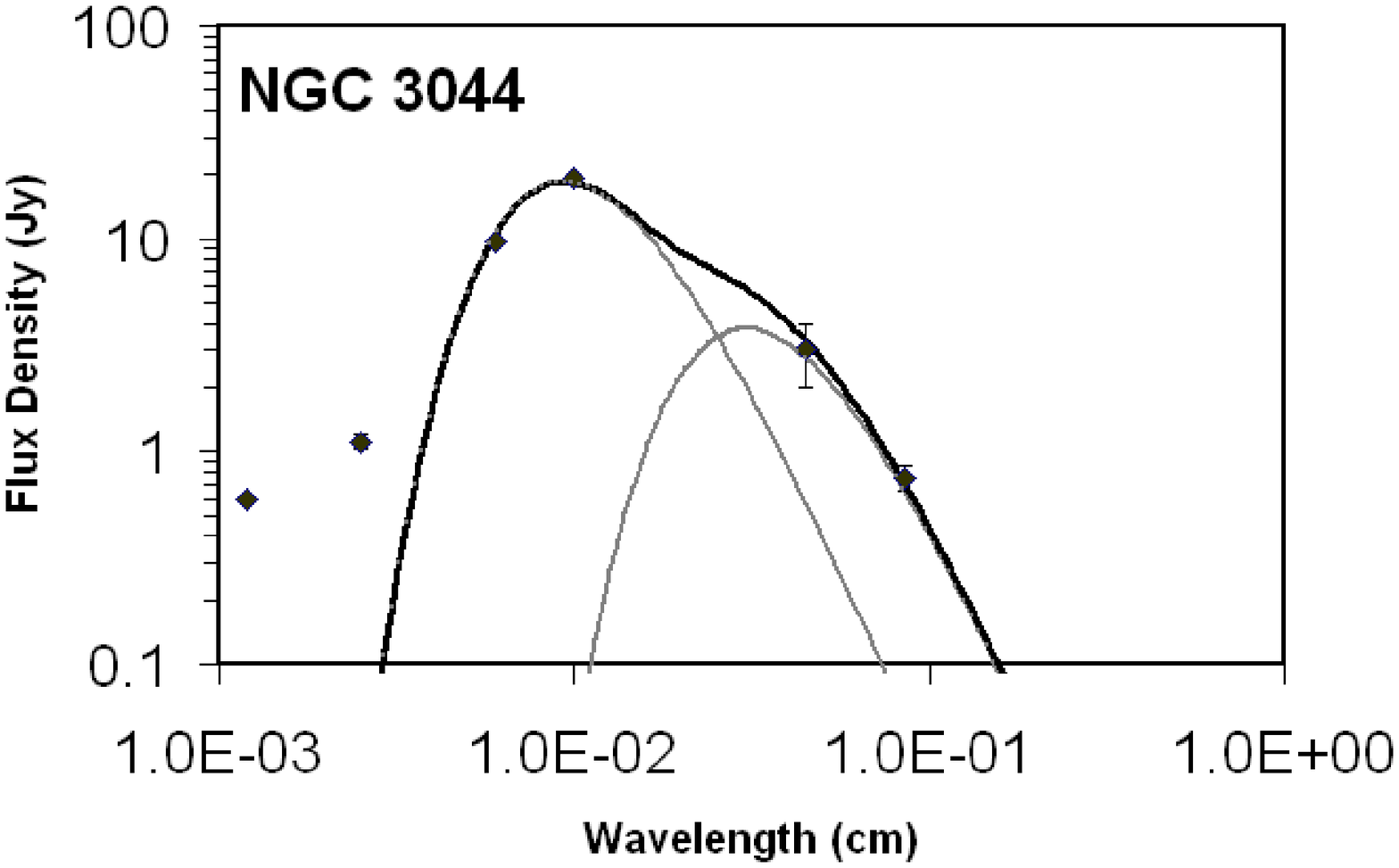}
\label{Tfita}
\end{minipage}%
\begin{minipage}{0.45\linewidth}
\centering
\includegraphics[width=7.5cm]{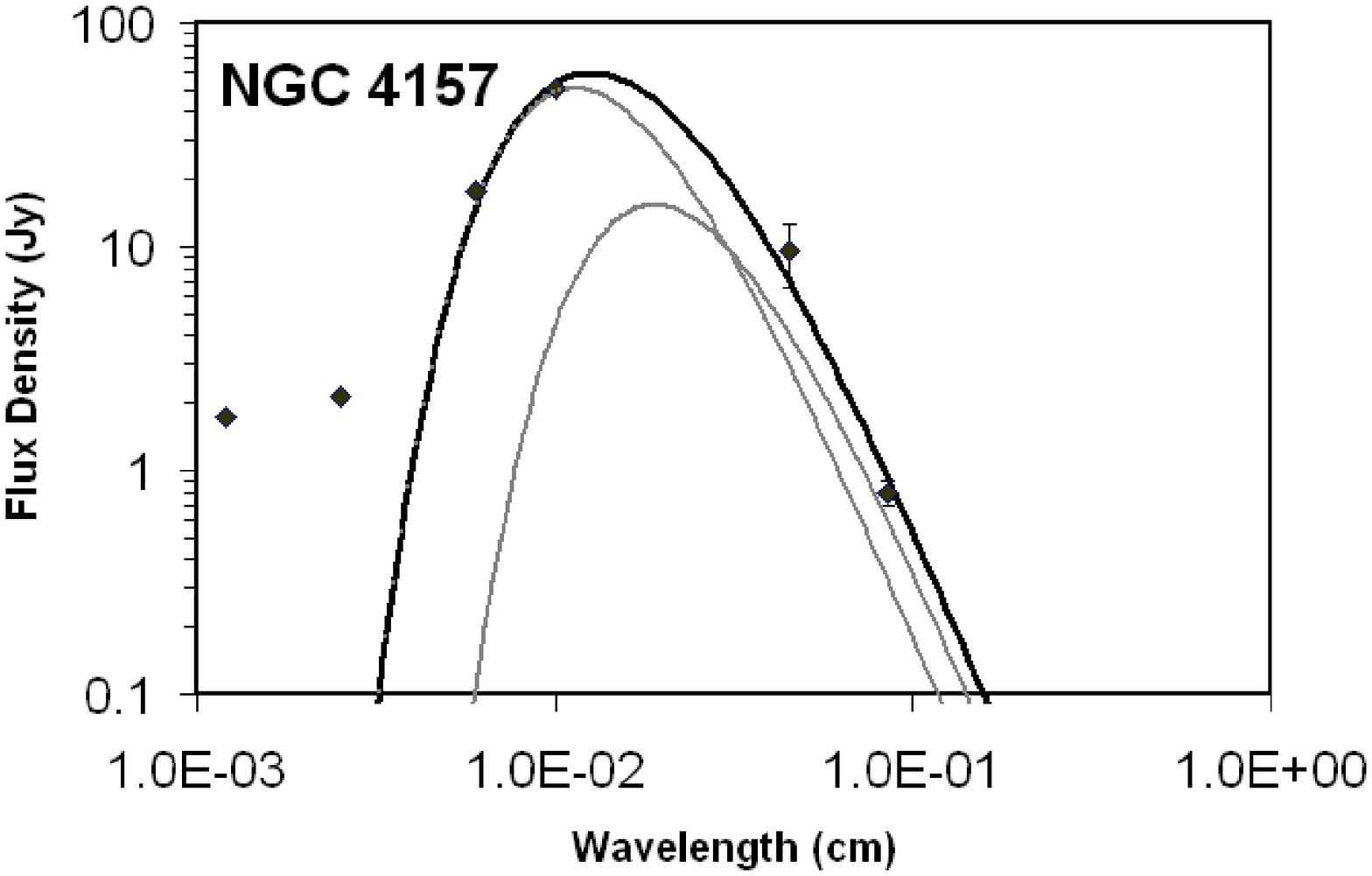}
\label{Tfigb}
\end{minipage}
\caption{The spectrum of NGC~3044 (left) and
NGC~4157 (right); the data points are shown with error bars.
A two-temperature fit with $\beta\,=\,2$ is shown for each.  Grey curves
show the warm and cold dust flux densities individually
 and the black curve shows the total fit
(see text).
\label{spectra_fig}}
\end{figure*}

The IR/sub-mm spectra of NGC~3044 and NGC~4157, which include
the IRAS flux densities (Table~\ref{tab:galaxy_parameters}) as well as the sub-mm data
(Table~\ref{properties_table}), are plotted in Fig.~\ref{spectra_fig}.

To determine the dust temperature, we consider only
those `classical' grains that are of sufficient size to reach an equilibrium
temperature, rather than very small grains (VSGs) which are 
stochastically heated.  Since the 12 and 25 $\mu$m flux densities are considered
to be contaminated by VSGs 
\citep[e.g.][]{dup03,des90}
we exclude these points from our fits.
  For the remaining points, we assume that the dust
is optically thin 
\citep[see, e.g.][]{mar97,eal96,ste00}
and that the data can be described by 
a modified black body curve of the form
\begin{equation}\label{eqn:modified}
S_\nu\,=\,\Omega_d\,(\nu/\nu_0)^\beta\,B_\nu({\rm T}_d)
\end{equation}
 where
$S_\nu$ is the flux density at frequency, $\nu$, $B_\nu({\rm T}_d)$ is
the Planck function for dust at
temperature, T$_d$, $\Omega_d$ is the solid angle subtended
by all dust (proportional to the number of dust grains, $N_d$, in the optically
thin limit),
$\nu_0$ is the frequency above which the dust becomes
optically thick, and  $\beta$ is the emissivity
index which we 
take to lie in the range, 1.5 to 2, as found from other
studies of our own and other galaxies \citep[e.g.][]{mas95,alt98,ste00}
The quantity, $(\nu/\nu_0)^\beta$ describes the
optical depth.  However, 
we could not fit the data, within errors bars,
with a single-temperature model.

Although it is most realistic for the dust to
have a spectrum of temperatures, we do not have sufficient data to test for
such a spectrum.  Instead, 
following \cite{dun01}, \cite{jam02}, \cite{alt00} 
and others, we fit a two-temperature (warm, denoted w, and cold, denoted c)
model to the data.  For identical dust grains that differ
only by temperature,   
the spectrum is fit according to,
\begin{equation}
\label{two_fit_eqn}
S_\nu\,=\,S_w\,+\,S_c\,=
\,K\,\nu^\beta\,\left[N_w\,B_\nu({\rm T}_w)\,+\,
N_c\,B_\nu({\rm T}_c)\right]
\end{equation}
where $N_w$ and $N_c$ represent the number of warm and cold dust grains,
respectively,  $S_w$ and $S_c$ represent the flux density contributions of warm and cold dust grains,
respectively, and   
$K$ is a constant that folds in the solid angle subtended by a dust grain
and the frequency at which the dust becomes optically thick.

Since we have 4 useable data points per galaxy, we can solve only for the
4 parameters:
$K\,N_w$, $K\,N_c$, ${\rm T}_w$, and
${\rm T}_c$.  From these, we list the physically meaningful parameters,
 $N_c/N_w$,  ${\rm T}_w$, and
${\rm T}_c$ in Table~\ref{properties_table} and show the fits 
in Fig.~\ref{spectra_fig}.  The uncertainties on the temperatures are associated
with 
the adopted range of $\beta$, i.e. $1.5\,\le\,\beta\,\le\,2.0$, which
dominates over other uncertainties.

The temperatures given in Table~\ref{properties_table} appear to be
typical of what has been found by other authors who have applied two-temperature
fits.  
For example, the ranges found by \cite{dun01} and \cite{jam02} 
for a large sample of galaxies 
were $28\,\le\,{\rm T}_w~({\rm K})\,\le\, 60$ and
$17\,\le\,{\rm T}_c~({\rm K})\,\le\, 32$.  
More recent data obtained from the Planck mission \citep{ade11} find
that, using two-temperature fits, a cold component (T$_c\,<\,20$ K) is
required in galaxies.  Our results for both NGC~3044 and NGC~4157 are consistent
with this.
We also find that
there is a greater range of dust temperature within 
NGC~3044 in comparison to NGC~4157. 
This appears to be consistent with the
slightly higher SFR and warmer $S_{60}/S_{100}$ colour (Table~\ref{tab:galaxy_parameters}) in NGC~3044, 
leading to higher values of
${\rm T}_w$, along with the fact that dust
emission has been detected to much higher galactocentric radii
 (compared to the optical
disk) in NGC~3044 as opposed to NGC~4157 (cf. Figs.~\ref{n3044resultsfig}d, 
\ref{n4157resultsfig}d) where lower ${\rm T}_c$ might be measured.


With the above assumptions, the fraction of cold to warm dust in the galaxy,
$N_c/N_w$ (equivalent to the mass fraction, $M_c/M_w$), has also been computed and is listed in
Table~\ref{properties_table}.    Previous values for a large sample of
galaxies \citep{dun01, jam02} fall in 
the rather large 
range, $0.6\,\le\,N_c/N_w\,\le\, 500$; with few exceptions,
the implication is that there is much more cold than warm dust in galaxies
and our results are again consistent with this conclusion.



For a two-temperature model, the dust mass can be found from
%
\begin{equation}
\label{dust_mass_eqn}
M_{d}\,=\,\frac{S_{\nu}\,D^2}{\kappa_{\nu}}
\left[\frac{N_w\,+\,N_c}{N_wB_\nu(T_w)\,+\,N_cB_\nu(T_c)}\right]\,
\end{equation}
where we take  $\nu\,=\,3.53\,\times\,10^{11}$ Hz (850 $\mu$m),
$S_{\nu}$ is the flux density at this frequency,
$D$ is the distance to the galaxy, and $\kappa_{\nu}$ is the dust mass opacity
coefficient at 850 $\mu$m. 
The value of $\kappa_{\nu}$ is uncertain since
$\kappa_\nu\,\propto\,\nu^\beta$.  We adopt a value of 
$\kappa_{\nu}\,=\,0.77$ cm$^2$ g$^{-1}$ for
consistency with other authors \citep{zhu07,vla05,dun01,dun00} 
and in agreement with \cite{jam02}.
The results are listed in Table~\ref{properties_table}.
  These masses fall within the range found
for previous galaxy samples observed at $\lambda\,850$ $\mu$m by \cite{jam02},
\cite{dun00} and also when $\lambda\,850$ $\mu$m data are included in 
SED fits \citep{wil09}.

There is considerably more dust in
NGC~3044 than NGC~4157.  Some of this difference may be attributed to sensitivity differences between
the two galaxies which did not enable detection of emission in NGC~4157 as far out in radius than
NGC~3044 (Sect.~\ref{sec:n4157_submm}).  
However, undetected flux at large galactocentric radii is unlikely to increase $M_d$ of NGC~4157
by the factor of 8 which would bring it into agreement with the dust mass in NGC~3044, even if
N$_c$/N$_w$ were increased to account for a possibly increasing fraction of colder dust at the larger
radii.
  There appears
to be a real difference in dust mass between the two galaxies.

The global gas/dust ratio for the two galaxies is given in 
Table~\ref{properties_table}
where the gas mass is the sum of $M_{HI}$ and $M_{H_2}$ in
Table~\ref{tab:galaxy_parameters} and the dust mass is given
in Table~\ref{properties_table}.  
The H$_2$ mass has been taken from early CO observations for which the
error bars are significant; nevertheless, the difference between
the two galaxies is again apparent, mainly because of the larger
dust content in NGC~3044, rather than significant differences between
gas masses.  (We have, in addition, resolved
CO(J=2-1) data for NGC~3044 from \cite{lee98} which we make use of
in the next section.)



\subsection{Major Axis Distributions of NGC~3044}

\begin{figure*}
\begin{minipage}{0.45\linewidth}
\centering
\includegraphics[width=7.5cm]{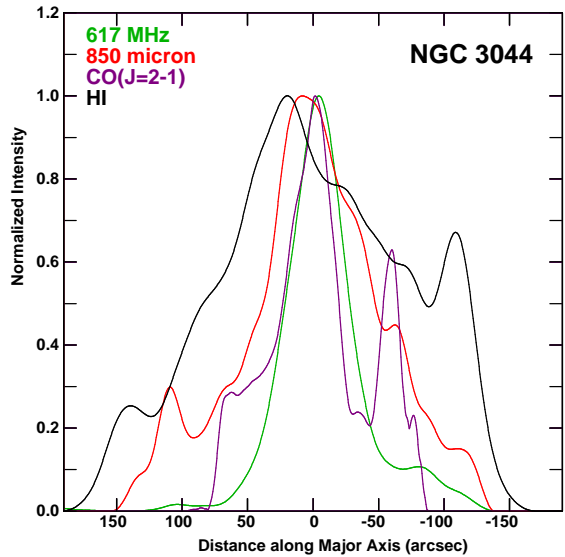}
\end{minipage}
\caption{Comparative major axis distributions of NGC~3044 for the
emission components as indicated at upper left,
normalized to their respective peak values.  The spatial resolution
is 21 arcsec for all distributions, the galaxy emission has been 
rotated by -23$^\circ$ so that the major axis lies along the x
axis and has been averaged over 21 arcsec in
the minor axis direction. 
Positive x values lie to the south-east along the major
axis.  Data sources and estimated errors
are:  617 MHz (this work, $\pm$0.004 normalized units),
850 $\mu$m (this work, $\pm$0.05 normalized units), 
CO(J=2-1) \citep[][20\%]{lee98}, HI
\citep[][$\pm$0.02 normalized units]{lee97}.  
\label{fig:majoraxis}}
\end{figure*}

For NGC~3044, we have spatially resolved data in a number of wavebands
for which we can compare the normalized
major axis distributions at a common spatial
resolution (21 arcsec or 2.2 kpc).  
Fig.~\ref{fig:majoraxis} shows
the $\lambda\,850$ $\mu$m slice in red and the 
617 MHz slice in green.  We show the $\lambda\,850$ $\mu$m data, rather than the
 $\lambda\,450$ $\mu$m data, since the S/N is higher for this data
set and also the data unambiguously represent the cold dust distribution
(Fig.~\ref{spectra_fig}).  A correlation between the cold dust and 
synchrotron emission is apparent, but there are significant differences as
well:  the synchrotron-emitting component is narrower in this
normalized plot and also has a smoother distribution.
We will return to this issue in Sect.~\ref{sec:correlations}.

In addition, we have obtained both HI and CO data for this galaxy from
\cite{lee97} and \cite{lee98}, respectively.  
The CO(J=2-1) data were obtained from
the JCMT 
 at 21 arcsec resolution and the data reduction
of this component follows the description in
\cite{lee01}; we take the CO(J=2-1) emission to 
represent the molecular gas distribution\footnote{Available CO(J=1-0) data are not
of sufficient spatial resolution \citep[see][]{sol88} for comparison.}.
These distributions are also plotted in Fig.~\ref{fig:majoraxis}.
The HI distribution is very broad in comparison to the molecular gas,
the latter being strongly centrally peaked with the exception of
a discrete peak at a projected radius of -60 arcsec
(i.e. to the northwest). 
The closest correlation is between
the synchrotron emission and the molecular gas distribution within
the central $\pm$ 50 arcsec, indicating the well-known close association
between molecular gas, star formation, and the subsequent synchrotron
radiation that is produced by supernovae. A plot of total gas distribution cannot
be formed since the point-by-point ratio, CO(J=2-1)/CO(J=1-0) is not available.

\subsection{617 MHz - $\lambda\,$850 $\mu$m Correlations}
\label{sec:correlations}

Following \cite{bra03} and, as indicated in Sect.~\ref{sec:fir-rad}, we have searched
for a correlation between the 617 MHz and  $\lambda\,$850 $\mu$m data within the two
galaxies; the latter waveband was chosen because it delinates the cold dust 
well (see Fig.~\ref{spectra_fig}) and also
has the highest S/N of our sub-mm data. We use the images of
Figs.~\ref{n3044resultsfig}b and d and \ref{n4157resultsfig}b and d, smoothed to
matching 15 arcsec spatial resolutions, sampled at 2 pixels per beam
(approximately Nyquist) and all images cut off at their
respective 3$\sigma$ noise levels.

\begin{figure*}
   \centering
   \includegraphics*[width=3in,height=3.5in]{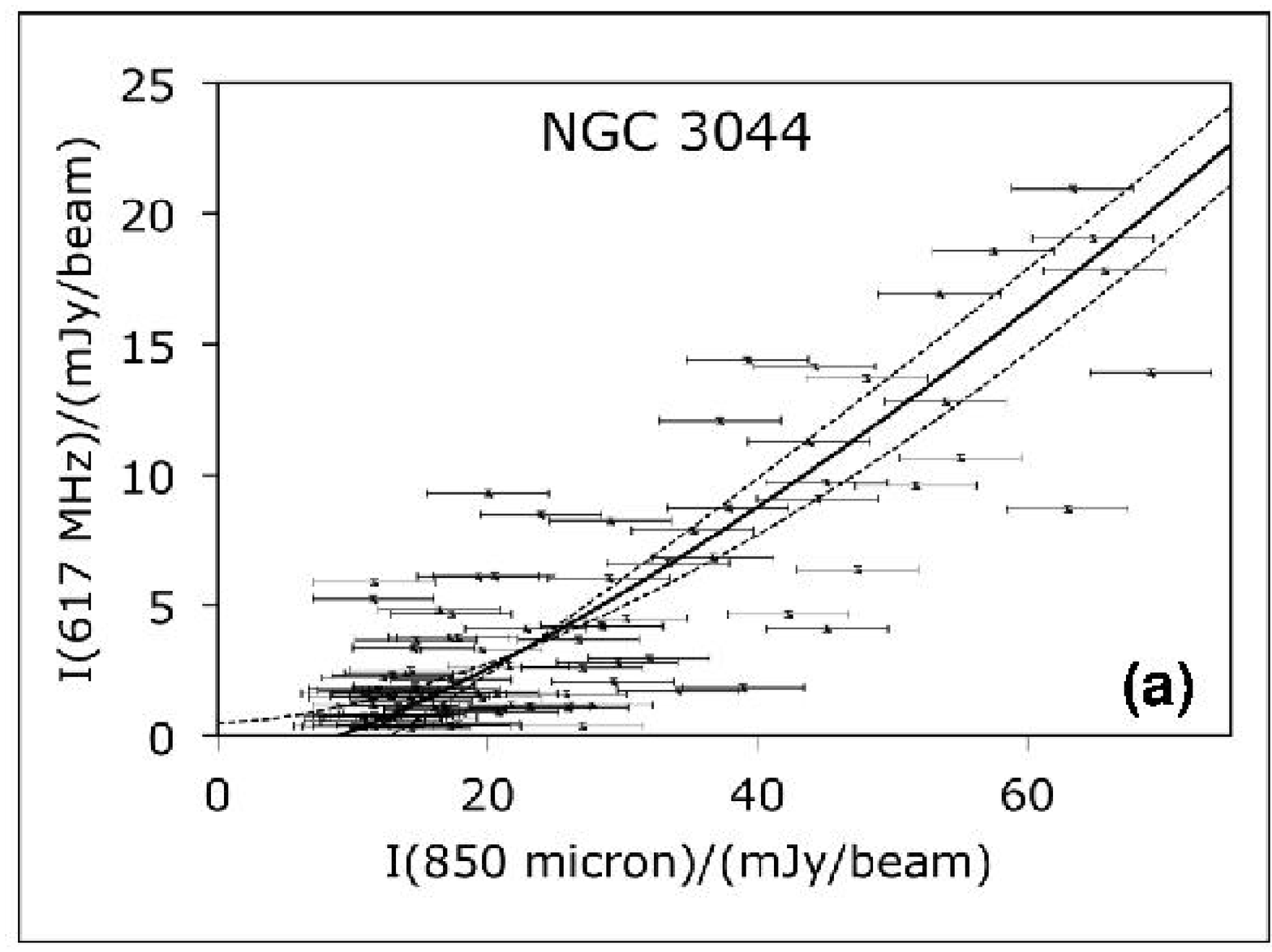}
   \hspace{0in}
   \includegraphics*[width=3in,height=3.5in]{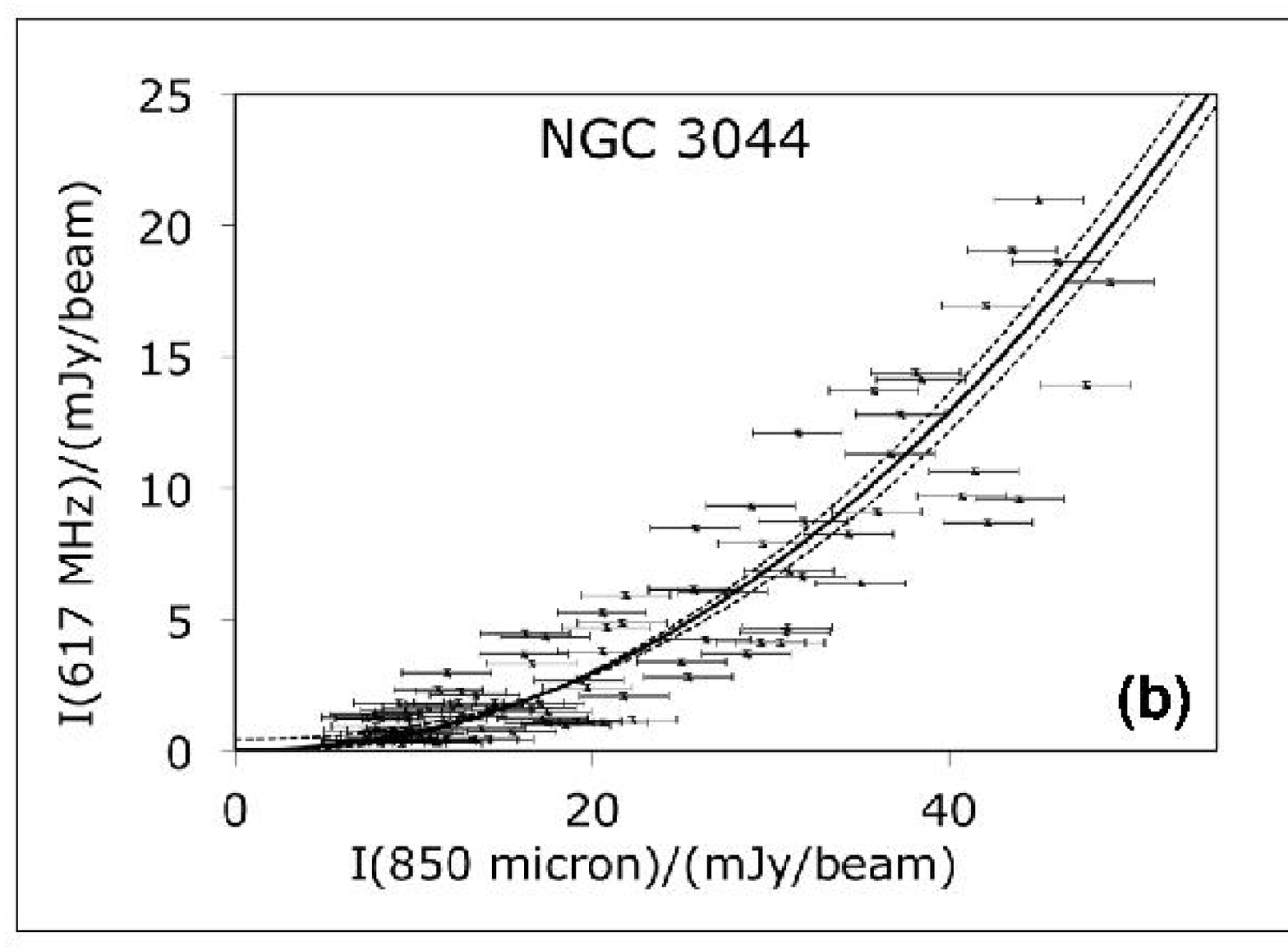}
   \hspace{0in}
   \includegraphics*[width=3in,height=3.5in]{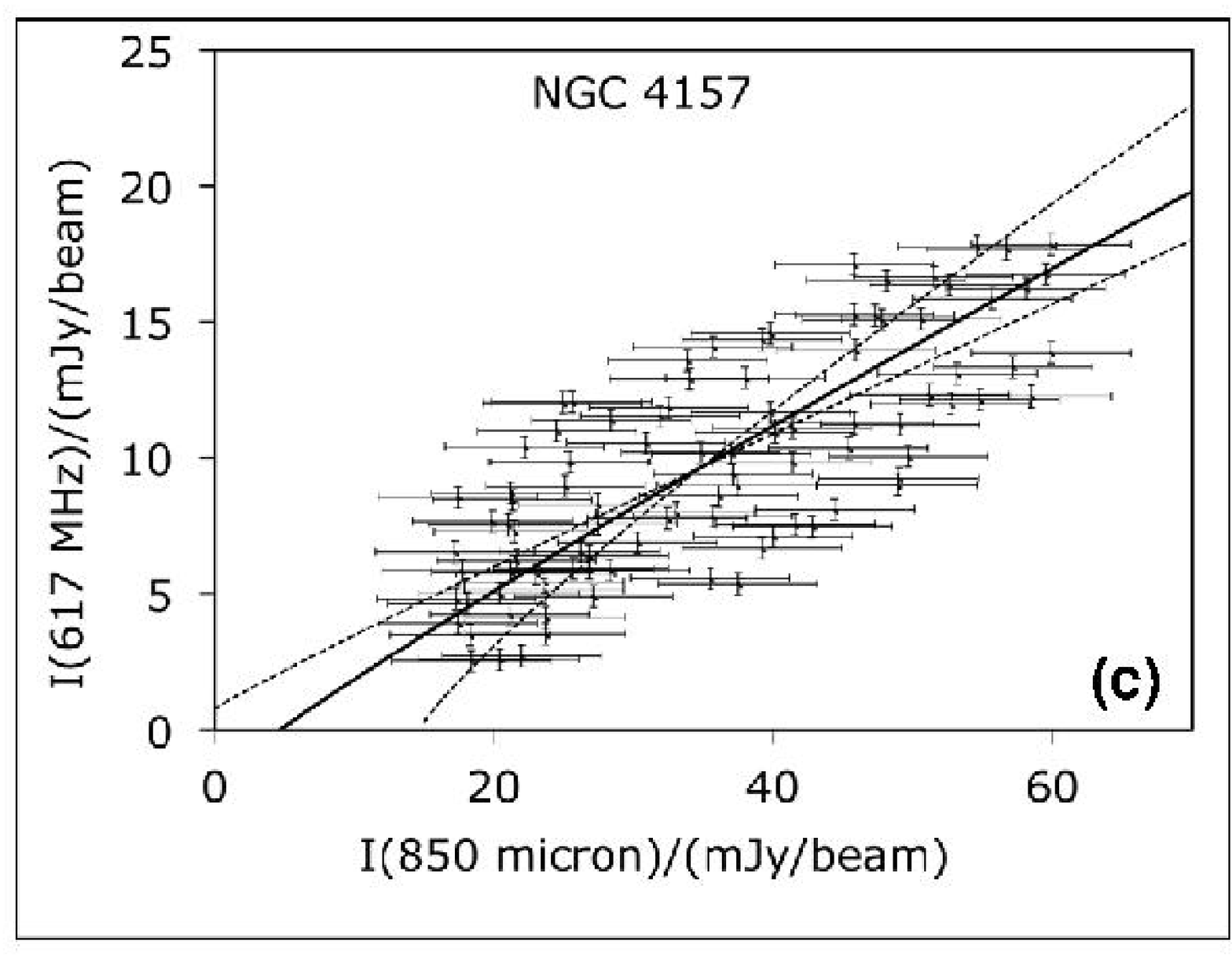}
   \hspace{0in}
   \includegraphics*[width=3in,height=3.5in]{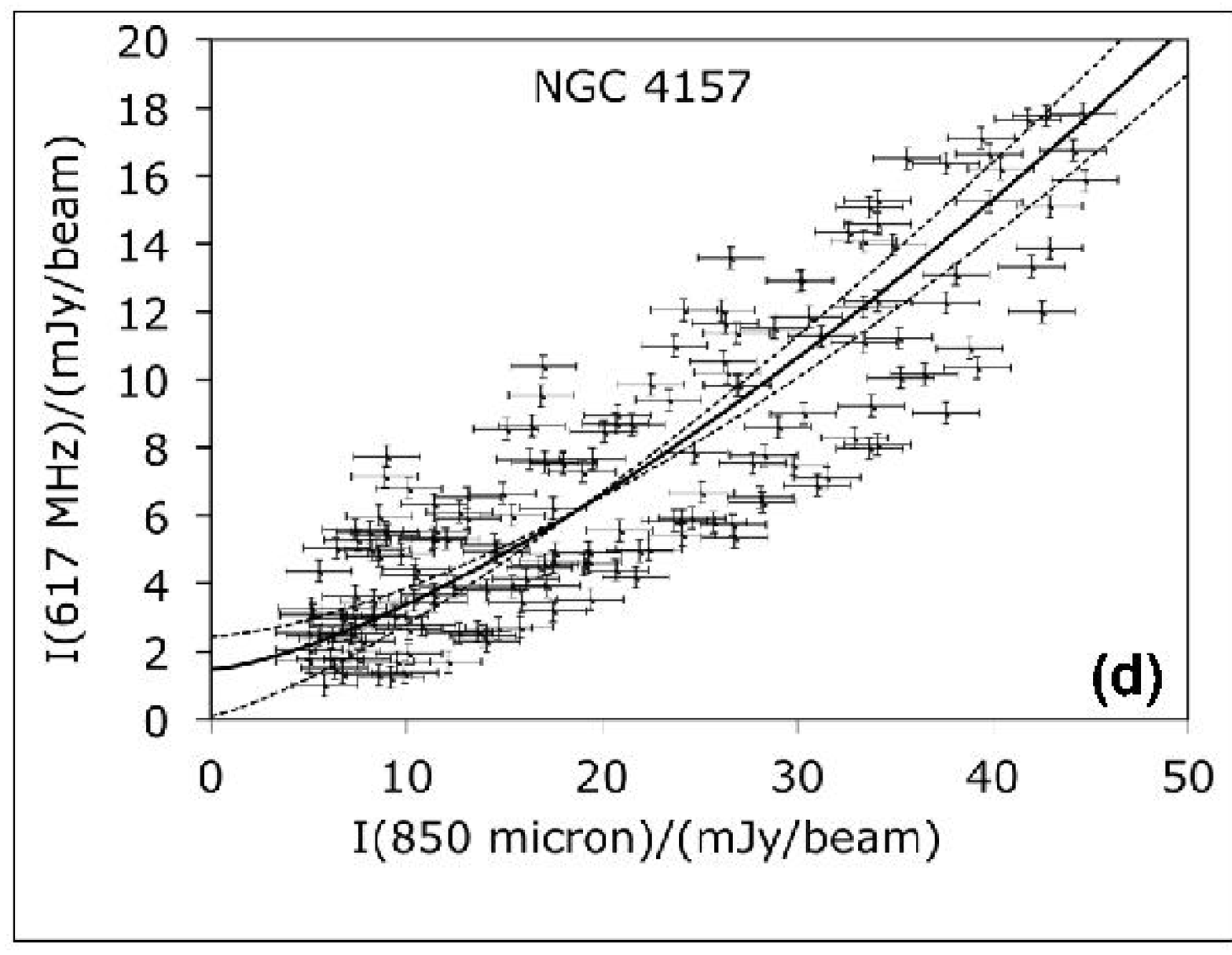}
   \caption{
Correlations between the $\lambda\,850$ $\mu$m and 617 MHz 
{specific intensities} for each
galaxy at 15 arcsec resolution and with each data set cut off below its respective
$3\,\sigma$ levels.  
Points with error bars are the measured values.  The two curved dotted lines are best fit
functions of the form, $y = ax^\alpha+b$, in the forward [y=f(x)] and reverse 
[x=f(y)] directions,
respectively (see Sect.~\ref{sec:correlations}),  
and the solid curves represent the bisector power law relations in each case.  
{\bf (a)} NGC~3044, with no smoothing.
{\bf (b)} NGC~3044, in which the $\lambda\,850$ $\mu$m image has been smoothed with a Gaussian
kernel of standard deviation, $\sigma_G\,=\,8$ arcsec. 
{\bf (c)} NGC~4157, with no smoothing.
{\bf (d)} NGC~4157, in which the $\lambda\,850$ $\mu$m image has been smoothed with a Gaussian
kernel of standard deviation, $\sigma_G\,=\,10$ arcsec.
}
\label{fig:correlations}
\end{figure*}

The results are shown in Figs.~\ref{fig:correlations}a and
c for NGC~3044 and NGC~4157, respectively, with the radio data on
the $y$ axis, as is customary.    
The 617 MHz and $\lambda\,$850 $\mu$m data
are clearly positively correlated within the two galaxies, the correlation coefficients
being 0.84 and 0.77 for NGC~3044 and NGC~4157, respectively. 
The scatter in the two plots exceeds the individual error bars on the points, a result which is also found
for the radio continuum - FIR relation, and the extent of the scatter appears to
be similar to that of the radio continuum - FIR relation, i.e. 
approximately a factor of 2 both within and between
galaxies \citep[e.g.][]{yun01,bel03,hoe98,hip03,mur06a}.

We then searched for the best relation between the two quantities
 using the least squares bisector approach
\citep{iso90} but as applied to potentially non-linear data. {Such an approach
allows for the presence of a non-zero intercept, should one exist.}
We use a Levenberg-Marquardt algorithm to search for a power law
of the form, $y\,=\,a\,x^\alpha\,+\,b$, which finds the best fit by minimizing the
weighted sum of the squared residuals in the $y$ coordinate. The fit is carried
out in both the 
 `forward' direction
(617 MHz data on the $y$ axis) as well as the `reverse' direction ($\lambda\,850~\mu$m data on
the $y$ axis).  The results of these two best fits are shown as dashed curves in
Fig.~\ref{fig:correlations}a and c.  The best fit is then considered to be the bisector
of the 
forward and reverse fits, where $\alpha$ is the average of the two fitted power law indices,
the bisector passes through the crossing point of the two fits, and the bisector is
placed such that areas between it and the forward and reverse fits are minimized. The
bisectors are shown as solid curves in Fig.~\ref{fig:correlations}a and c and their parameters
are listed in 
 Table~\ref{tab:fir_rad}. 

{The results of Table~\ref{tab:fir_rad} do not change significantly if the lower cutoff is varied or if the
pixel sampling is varied. 
For example, we have tried a lower cutoff of 2$\sigma$ and also
one pixel-per-beam averaging, yielding consistent results. The uncertainties 
 are given for $\alpha$, which  is the most
important fitted parameter (see Sect.~\ref{sec:discussion}), determined from the larger of the
formal error of the fit and some variation that results from systematic trials with
different input parameters as starting points, and finally adjusting to one pixel/beam sampling.
In each
case, a power law fit improved the result over a simple linear fit (i.e. $\chi^2_{r}$ is lower),
although, for both galaxies,
the uncertainty on the power marginally encompasses the linear case
($\alpha\,=\,$1.4$\,\pm\,$0.5 for NGC~3044 and
0.91$\,\pm\,$0.08 for NGC~4157).}

\begin{table*}
 \centering
 \begin{minipage}{170mm}
  \caption{Parameters of the 617 MHz - $\lambda\,850$ $\mu$m  
Correlations\label{tab:fir_rad}}
  \begin{tabular}{@{}cccccccccc@{}}
  \hline
                     \multicolumn{5}{c}{NGC~3044\footnote{An equation
of the form, $S_{617}\,=\,a\,S_{850}^\alpha\,+\,b$ has been applied,
where  $S_{617}$ and
$S_{850}$ are the 617 MHz and 850 $\mu$m flux densities, respectively, in
 mJy beam$^{-1}$.}} & 
\multicolumn{5}{c}{NGC~4157$^a$} \\
  $\sigma_G$\footnote{Standard deviation of the Gaussian smoothing kernel
applied to the $\lambda\,850~\mu$m data. Blank means
 no smoothing.}& 
$a$\footnote{Fitted parameters as defined in $a$ above.} 
& $\alpha$$^c$ & $b$$^c$  &$\chi^2_{r}\,$\footnote{Reduced $\chi^2$, i.e. 
weighted sum of the squared residuals in the $y$ coordinate, normalized by the number
of degrees of freedom.}
& $\sigma_G$$^b$ & $a$$^c$ &$\alpha$$^c$  &  $b$$^c$ &$\chi^2_{r}\,$$^d$ \\
\hline
  & 0.06  & 1.4 $\,\pm\,$0.5&  -1.3 & 7.55 &  & 0.046 & 0.91$\,\pm\,$0.08 & -1.9 & 7.34\\
 $8^{\prime\prime}$ (0.84 kpc)  &  0.005 $\pm$ 0.003 & 2.1$\,\pm\,$0.3 & 0.04 $\pm$ 0.2 & 3.06 &  
$10^{\prime\prime}$ (0.63 kpc) & 0.07 $\pm$ 0.05 & 1.4$\,\pm\,$0.3 & 1.5 $\pm$ 0.5  & 4.9\\
\hline
\end{tabular}
\end{minipage}
\end{table*}


As indicated in Sect.~\ref{sec:fir-rad}, the FIR-radio continuum relation has been shown to
improve when the FIR emission is smoothed.  This process effectively `forces' the UV photon
dust optical depth 
to more closely match the synchrotron diffusion length. 
Following \cite{mur08}, we apply a sequence of smoothing kernels to the same
$\lambda\,850~\mu$m maps used above
with consecutively increasing `smearing scale-lengths'.  A 
 $\lambda\,850~\mu$m/617 MHz ratio map is then formed, normalized, and the residuals
compared to see which, if any, improve upon the relations plotted in 
Figs.~\ref{fig:correlations}a and c.  We tried both exponential smearing kernels
($e^{r/{l_e}}$) where $r$ is the distance from each point and $l_e$ is the exponential
scale length (both in the plane of the sky) as well as gaussian smearing kernels
($e^{r^2/2{{\sigma_{G}}^2}}$), where $\sigma_G$ is the standard deviation
 of the gaussian. In both
cases, a sequence of scales from 1 to 25 arcsec in 1 arcsec steps was tried.  Like
\cite{mur08}, we find that the residuals first decrease and then increase again as
kernel size increases,
 resulting in a clear scale for which the fit is best.  

\begin{figure}
\includegraphics*[width=7.75cm]{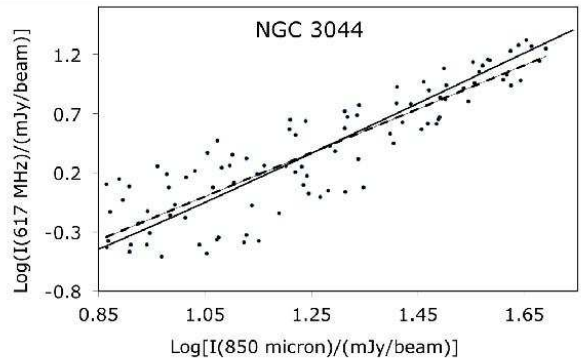}
  \caption{Logarithmic plot of the $\lambda\,850~\mu$m 617 MHz correlation for NGC~3044.  The data from
Fig.~\ref{fig:correlations}b have been replotted here without the error bars.  The solid curve is the
bisector from that figure and the dashed curve is the best linear fit in logarithmic space whose slope
is 1.9 $\pm$ 0.1.
}
\label{fig:logplot}
\end{figure}

The best fit results are given
in Table~\ref{tab:fir_rad} and are plotted in
Figs.~\ref{fig:correlations}b and d. There is no doubt that the fits are improved after
applying 
this smoothing and, moreover, {that the correlation becomes non-linear
($\alpha\,=\,$2.1$\,\pm\,$0.3 for NGC~3044 and
1.4$\,\pm\,$0.3 for NGC~4157).  For these cases with smoothing kernels,
 we also provide uncertainties on the
other fitted parameters, as described above. Note that the intercept for NGC~3044 
(Table~\ref{tab:fir_rad}) is consistent
with zero.  This is not the case for NGC~4157 although, for both galaxies, the forward and
reverse bisectors encompass the (0,0) point.  Since the intercept is zero for NGC~3044,
we have also provided a logarithmic plot for this galaxy which is shown in Fig.~\ref{fig:logplot}.
The best fit linear slope in this space with the displayed sampling
(1.9 $\pm$ 0.1) agrees with the non-linear fit of
Table~\ref{tab:fir_rad}.} 

We find that the Gaussian smoothing kernel produces a better fit than an exponential kernel.
It has been suggested that an exponential kernel is more likely to describe a situation in
which cosmic ray electrons escape from the disk whereas a Gaussian kernel
would describe a random walk \citep{bic90}. For an edge-on galaxy, it is not straightforward
to interpret the resulting best fit scale; however, since we are looking through a
long in-disk line of sight in our two galaxies, we would expect that the in-disk (random
walk) component should dominate, consistent with our
better Gaussian fit.  For more face-on galaxies, 
\cite{mur06a} found that exponentials provided better fits and 
\cite{mar98}, with earlier data, found that both Gaussians and exponentials were
equally acceptable.

The non-linear relationship between the 617 MHz and $\lambda\,850$ $\mu$m emission explains
the fact that the major axis slices for these two bands, although roughly correlated, do not follow each
other particularly well in
Fig.~\ref{fig:majoraxis} (red and green curves).
However, when the functional relationship shown in Fig.~\ref{fig:correlations}b is applied
to the $\lambda\,850~\mu$m emission and renormalized, the widths of the two distributions
(not shown) match each other
to well within a beam size. (Note that squaring the $\lambda\,850~\mu$m emission will {\it narrow} its
distribution, bringing it in alignment with the 617 MHz emission after normalization.)
These results will be discussed further in Sect.~\ref{sec:discussion}.

\section{Discussion}
\label{sec:discussion}

\subsection{High-Latitude Emission}

\subsubsection{NGC~3044}
\label{sec:n3044-discuss}
The presence of a multi-phase gaseous halo in NGC~3044 has been known for some time
\citep{hum89,con90,col96,irw99,irw00,col00,tul00,mil03,tul06,lee97}.  
See \citet{col00}, for example,
for measurements of the HI and eDIG scale heights in this galaxy. 
Our
new observations are the first at 617 MHz and $\lambda\,850~\mu$m and reveal several
new high latitude features in these two wavebands
(Fig.~\ref{n3044resultsfig} and Sect.~\ref{sec:radcont_images}). 

\begin{figure}
\includegraphics*[width=3.2in]{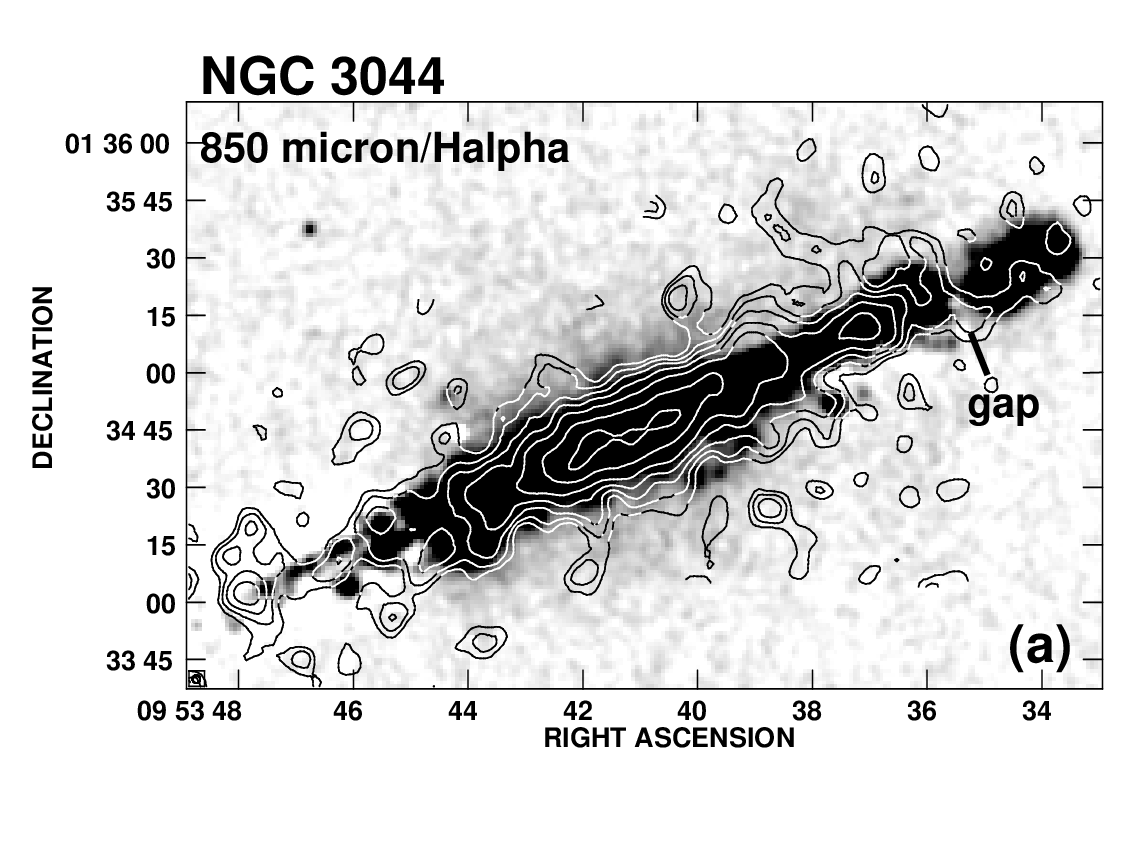}
\vskip -0.3truein
\includegraphics*[width=3.2in]{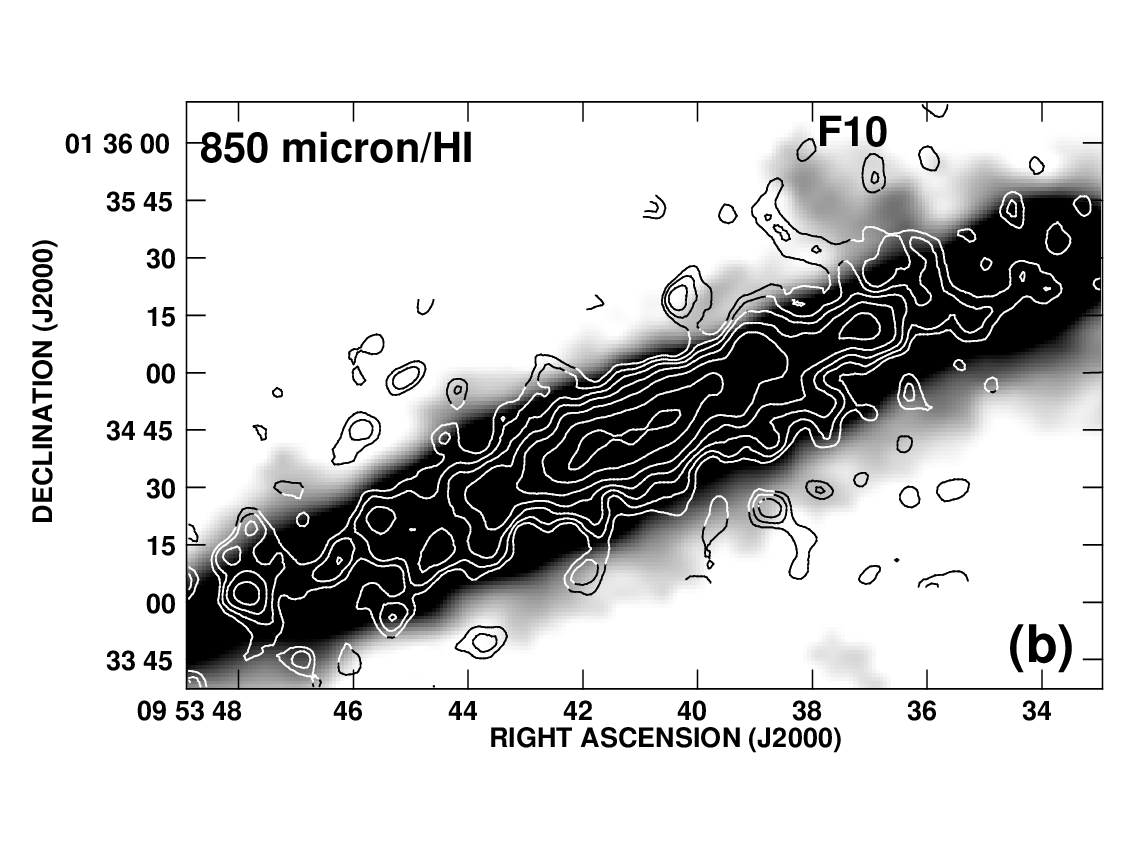}
\vskip -0.3truein
\includegraphics*[width=3.2in]{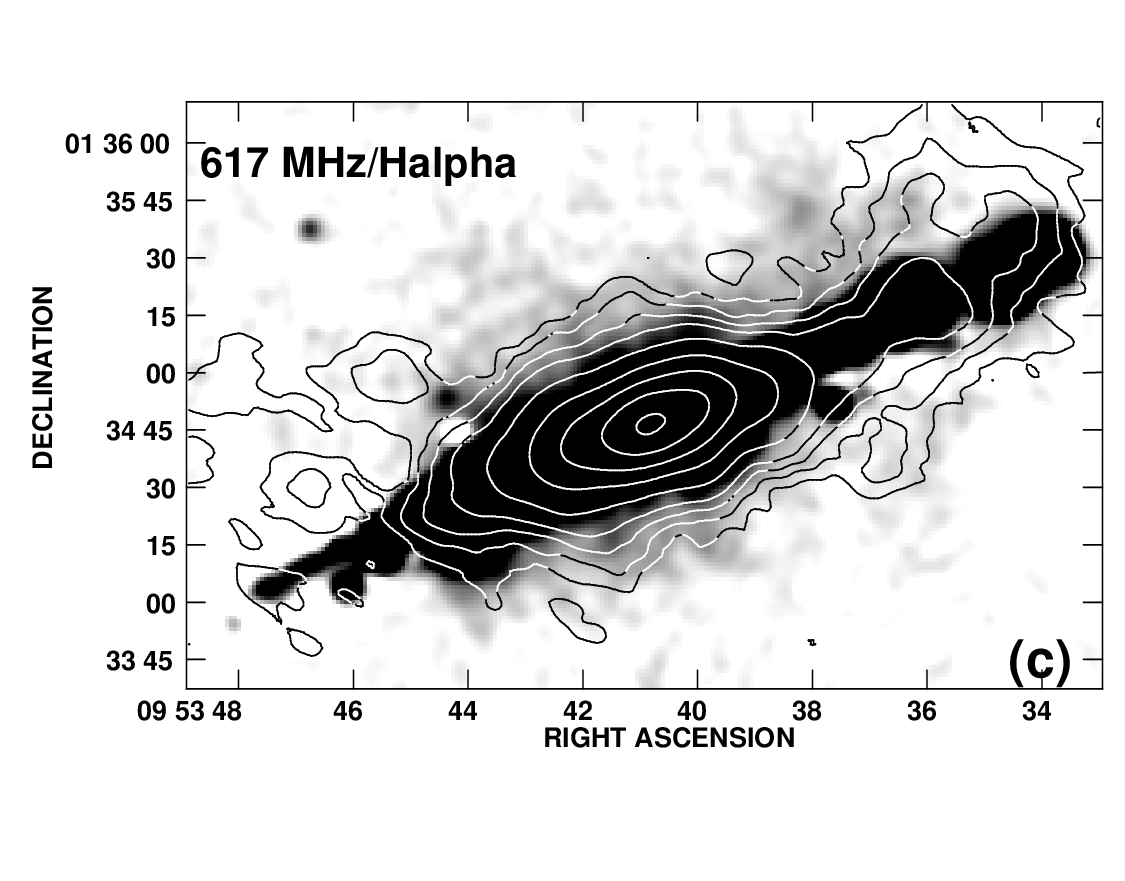}
\caption{Selected overlays for NGC~3044 showing extraplanar emission.
{\bf(a)} $\lambda\,850~\mu$m contours as in Fig.~\ref{n3044resultsfig}d, over a greyscale
H$\alpha$ image from \citet{col00}. The H$\alpha$ `gap' in the major axis
that is discussed in Sect.~\ref{sec:n3044-discuss} is
labelled.
{\bf (b)} $\lambda\,850~\mu$m contours as in Fig.~\ref{n3044resultsfig}d, over a greyscale
HI total intensity map from \citet{lee97}, where F10 denotes an HI extension identified as part of
an expanding shell in the latter reference.
{\bf (c)} 617 MHz contours as in Fig.~\ref{n3044resultsfig}b over a greyscale
 H$\alpha$ image from \citet{col00}
smoothed to 5 arcsec resolution.
 }
\label{fig:n3044_overlays}
\end{figure}

These disturbances are better seen in several
 selected overlays shown in Fig.~\ref{fig:n3044_overlays}; the greyscales show H$\alpha$
emission, HI total intensity emission, and smoothed H$\alpha$ emission in frames a, b, and
c, respectively.
At a location of about 1.5 arcmin from the nucleus along the
north-west major axis there is a gap in the H$\alpha$ emission in the disk
(Fig.~\ref{fig:n3044_overlays}a).  At this position is an HI supershell,
part of which is visible as the high latitude feature, F10 
(Fig.~\ref{fig:n3044_overlays}b), so labelled by \citet{lee97}.  The narrow vertical
dust feature observed at $\lambda\,850~\mu$m, although at a low S/N, occurs adjacent to F10 and the
double-pronged 617 MHz feature is also at this position.  Apparently, there is
disturbance in the disk at this location which is affecting all of the
components displayed and that disturbance has excavated an ionized gas cavity in
the disk.  
This is not the first time that high latitude features
have been associated with a gap or absence of H$\alpha$ emission in the underlying
disk; see \citet{lee01} for another prominent example in NGC~5775.


On the whole, there is global correspondence between the radio continuum and H$\alpha$ emission
as seen in Fig.~\ref{fig:n3044_overlays}c). This correspondence,
which has been pointed out by many authors, including \citet{col00} for NGC~3044 itself,
 can be understood
from the fact that both components are associated with massive star formation.  

What is more 
unusual, as pointed out in Sect.~\ref{sec:radcont_images}, are the extraplanar features
which appear to originate at the two ends of the major axis.  In particular, at the south-east
end, the emission is truncated approximately where the 
 H$\alpha$ emission also
narrows abruptly (Fig.~\ref{fig:n3044_overlays}c).  We then see a series of disconnected
emission features in {\it both} the GMRT 617 MHz and VLA 20 cm images
(Figs.~\ref{n3044resultsfig}b and \ref{n3044vlafig}b, respectively), including the
feature at RA $\approx$ 9$^{\rm h}$ 53$^{\rm m}$ 51$^{\rm s}$,
DEC $\approx$ 01$^\circ$ 35$^\prime$ 45$^{\prime\prime}$ which we have labelled 
the non-thermal cloud (NTC) in  Fig.~\ref{n3044resultsfig}b. The feature centers at approximately
2 arcmin (13 kpc)
from the galaxy's major axis.

\begin{figure}
\includegraphics*[width=8.3cm]{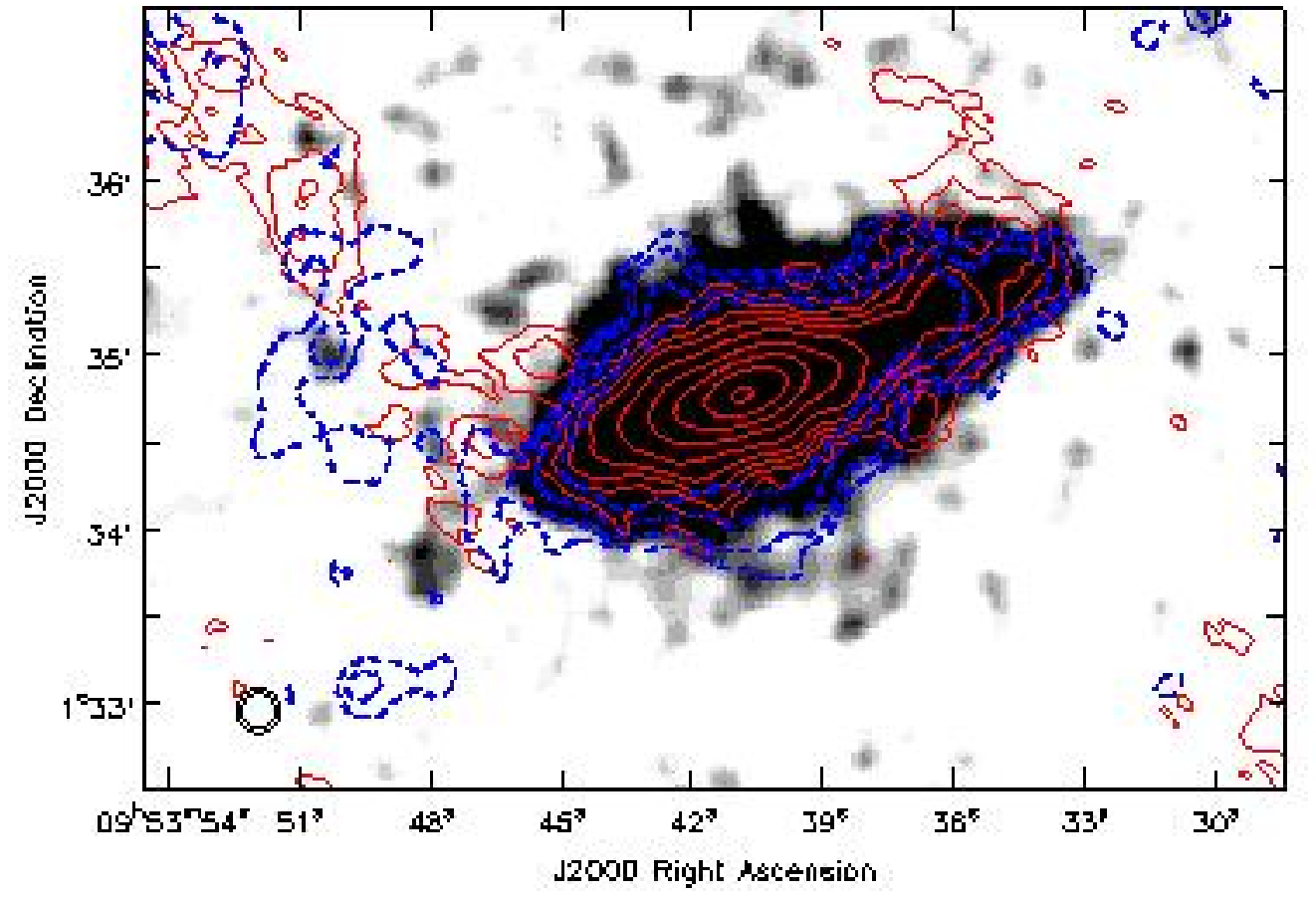}
\caption{The 617 MHz emission (red contours) from Fig.~\ref{n4157resultsfig}b,
and the $\lambda\,20$ cm emission (blue dashed contours) from Fig.~\ref{n4157vlafig}b,
are shown overlaid on a greyscale of the the WISE $\lambda\,22$ $\mu$m emission.  The WISE
image has been cut off at the 2$\sigma$ level and is shown so as to
enhance the faint broad-scale emission.
 }
\label{fig:n3044_wise}
\end{figure}

In Fig.~\ref{fig:n3044_wise} we show the 617 MHz (red contours) and $\lambda\,20$ cm 
(blue dashed contours) emission 
of NGC~3044 superimposed on 
a greyscale image of the $\lambda\,22$ $\mu$m emission taken from the Wide-field Infrared Survey Explorer (WISE)
all sky survey \citep{wri10,cut12}, the latter enhanced to show low intensities that are $>$ 2$\sigma$ in brightness.
The $\lambda\,22$ $\mu$m emission is an indicator of the presence
of warm dust (of order 130 K for classical grains in thermal equilibrium). This map shows that the
halo of NGC~3044 is even more extensive than previously known and shows considerable substructure,
including vertical extensions and arcs as well as `disconnected' features towards the north. 
It is beyond the scope of this paper to discuss the WISE results in detail; however, when considered together
with the two independent radio continuum maps (GMRT and VLA), the results give credence to the reality of the northwards
extensions that begin on the south-east end of the major axis.  Note that, at low
S/N, sensitivity to spatial scales (which depends on an array's uv coverage) is as important as S/N in
determining what emission features will be detected.  Consequently, the GMRT and VLA emission, while overlapping,
do not align perfectly and should not be used to determine spectral indices at these low emission levels.
Put together, though, the truncation of the radio continuum major axis where the northwards extension begins, 
the abrupt narrowing of
the H$\alpha$ emission at this position, and emission extending towards the north that is visible in GMRT
617 MHz, the VLA $\lambda\,20$ cm map, and the WISE $\lambda\,22$ $\mu$m map, suggest that a disruption
has occurred at the south-east end of the major axis in this galaxy and expelled material to the north of the
plane.


Bearing in mind the low S/N of the NTC, if we nevertheless take it at face value, we can
estimate some physical parameters.
Excluding the smaller features closer to the
major axis, the NTC, as shown in Fig.~\ref{n3044resultsfig}b, has a flux density of 3 mJy,
or a spectral power of $P_{617~MHz}\, =\, 1.7\,\times\,10^{20}$ W/Hz.  By comparison,
the flux density of the Galactic supernova remnant, Cas A, is $S_{617}(Cas~A)\,=\,3425.4$ Jy
for epoch 2005.5
\citep{vin07},
corresponding to  $P_{617~MHz}\, =\, 4.7\,\times\,10^{18}$ W/Hz at its distance of 3.4 kpc
\citep{ham08}.  Thus, the radio power of the NTC alone corresponds to 36 equivalent supernova remnants.
If we include features closer to the disk, the value approximately doubles.  
The environment of the NTC differs substantially from that of Cas A; nevertheless this comparison
suggests that a massive star-forming region in the disk could indeed have been
responsible for the observed emission from the point of view of energetics.

Regarding timescales, from the mean magnetic field strength ($\overline{B}\,=\,5.8$ $\mu$G) found throughout the disk and
the observed halo of NGC~3044 \citep{irw99}, the mean cosmic ray electron lifetime is
$\tau_{CR}\,=\,6.5\,\times\,10^7$ yr
(see Sect.~\ref{sec:correlation-discuss}). However, the magnetic field was found to vary from 
3.5 to 8.9 $\mu$G and it is
likely that CRs far from the disk would be associated with magnetic fields that are
at the low end of this range.  Consequently,
a lifetime of $\tau_{CR}\,=\,1.4\,\times\,10^8$ yr ($\tau_{CR}\,\propto\,B^{-3/2}$) may be more appropriate.
 For a distance of
13 kpc, then, the outflow velocity would be 91 km s$^{-1}$ which is much less than the 300 km s$^{-1}$ outflow 
velocity observed in NGC~253 by \cite{hee09}, for comparison.

The above calculations are order of magnitude only and are
meant to illustrate the feasibility of non-thermal emission far from the
disk of NGC~3044 having originated from activity within the disk.  We require more information
on the magnetic field distribution and strength to make more definitive statements.

As noted earlier, the 617 MHz emission in NGC~3044 shows curvature towards the north on both ends of the major
axis of NGC~3044 which sometimes suggests that ram pressure stripping could be occurring as
the galaxy
passes through an intergalactic medium (IGM) in a southerly direction.
However, there is no such evidence in the HI emission and, in fact, HI extensions are observed 
on both the north and south sides of the major axis.
In addition, the isovelocity HI contours show a slight curvature
which, if caused by motion through an IGM, would imply a motion to the north-east
rather than the south-west
\citep[see Fig. 4 of][]{lee97}.
This suggests that the observed 617 MHz extensions
at the ends of the major axis and the disconnected northern features
may have been produced from internal activity which is asymmetrically placed
with respect to gas distribution (or magnetic field distribution) in the galaxy.
  \citet{lee97} also argue for
an internal origin for the HI supershells and the energetics (above) are consistent with
this.  With distributed star formation in the galaxy, it is reasonable to expect that
outflows will more effectively eject material far from the plane at large
galactocentric radii where the internal
disk density is lower.

\subsubsection{NGC~4157}
\label{sec:n4157-discuss}

NGC~4157 is also known to exhibit high latitude emission \citep{irw99,ken09} but fewer observations
of this galaxy have been carried out in comparison to NGC~3044.
As seen in Figs.~\ref{n4157resultsfig}c and d, we do not have sufficient S/N for a clear
detection of high latitude dust in NGC~4157.  The radio continuum emission 
(Figs.~\ref{n4157resultsfig}a and b) shows some
extensions that are roughly at the locations of those seen in the VLA 20 cm map
(Fig.~\ref{n4157vlafig}).  A new feature is seen in the 617 MHz map (Fig.~\ref{n4157resultsfig}b)
in the form of a radio continuum extension beginning at the far north-eastern end of the major
axis and extending south, as described in Sect.~\ref{sec:n4157_radio}.

\begin{figure}
\includegraphics*[width=8cm,scale=1.0]{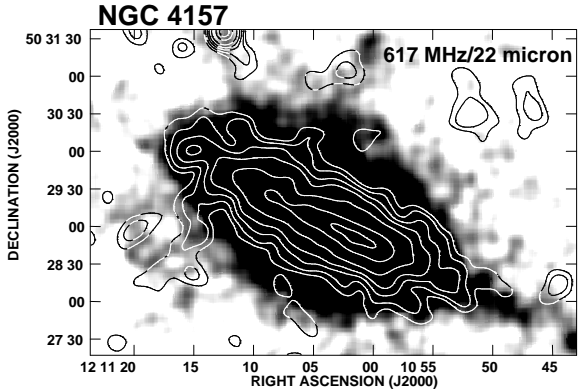}
  \caption{Contours of GMRT 617 MHz emission as in Fig.~\ref{n4157resultsfig}b over a greyscale
WISE image at $\lambda\,22~\mu$m.  The WISE image
(spatial resolution of 12 arcsec) has been cut off at 2$\sigma$ and is shown so as to enhance
the faint broad-scale
emission.
}
\label{fig:n4157_overlay}
\end{figure}

This southern extension can be better seen in Fig.~\ref{fig:n4157_overlay} where it is overlaid onto
a greyscale image of the $\lambda\,22$ $\mu$m WISE emission.
This plot 
shows that warm dust exists in a halo that extends up to
1.3 arcmin (4.9 kpc) from the plane.  
The HI in this galaxy extends to roughly the same {\it z} distance \cite{ken09} (although HI also extends
much farther in the radial direction along the major axis).  The dust, therefore, may be a component of
the high-latitude HI in the same way that it normally is within a galaxy disk.  

For the purposes of
our 617 MHz comparison, we note that the southern 617 MHz extension coincides with discrete
dust emission in roughly the same direction which gives weight to our assertion that the extension is
real.  
Indeed, some of the 617 MHz features on the north side of the disk also have dust counterparts.
Although it isn't clear whether these features occupy the same volume, the evidence continues to
mount that all constituents of the ISM are represented in galaxy halos.  See \citet{lee01} for
an earlier example.

\subsubsection{The Non-thermal - Cold Dust Relation in NGC~3044}
\label{sec:correlation-discuss}

As can be seen from Fig.~\ref{spectra_fig}, the $\lambda\,850~\mu$m emission in NGC~4157 (as modeled)
still includes a contribution from the warm dust component whereas the emission of NGC~3044 is strongly
dominated by only cold dust at that wavelength.  In addition, the S/N of the NGC~3044 data
is higher than that of NGC~4157.  We therefore consider only NGC~3044 in our discussion of the 617 MHz -
$\lambda\,850~\mu$m relation which we take to represent a relationship between synchrotron emission
and cold dust (T$_c\,=\,9.5\,\pm\,1.5$ K, Table~\ref{properties_table}).

Firstly, we note that the relationship between these two components is significantly improved when a smearing
kernel is introduced to the sub-mm data (Fig.~\ref{fig:correlations}), consistent with earlier results
\citep{mur06a,mur06b,mur08,mur09}.
 This kernel `corrects' for differences between
the photon mean free path to dust absorption and the CR electron diffusion length.  We find, not only that the
correlation improves, but also that it becomes clearly non-linear such that 
$S_{617~MHz}\,\propto\,S_{850~\mu{\rm m}}^{2.1\,\pm\,0.2}$; that is, the relationship is roughly consistent with
 a luminosity relation of $L_{synch}\,\propto\,L_{cold~dust}^{2.1}$.  By comparison, \citet{pie03} found
 $L_{1.4~GHz}\,\propto\,L_{cold~dust}^{1.13}$ and \citet{bra03} found a roughly linear 
617 MHz - $\lambda\,850~\mu$m relation for NGC~5775.
Neither of these authors applied a smearing kernal and, as  Fig.~\ref{fig:correlations} 
illustrates, the correlation becomes much flatter ($S_{617~MHz}\,\propto\,S_{850~\mu{\rm m}}^{1.4\,\pm\,0.3}$,
Table~\ref{properties_table}), when a smoothing kernel is ignored. 

\citet{mur06b} have shown that galaxies with higher SFRs per unit area have lower kernel scale lengths.
They interpret this result to imply that galaxies with high SFRs contain, on average, younger SF regions
from which CR electrons have not had sufficient time to diffuse to large scales. When SF has not been
so recent, by contrast,
the difference between the radio continuum scale 
and the dust scale diminishes.  

As indicated 
in Sect.~\ref{sec:correlations},
it is not straightforward to interpret the meaning of the kernel size for an edge-on galaxy in which the line of
sight that is being probed varies with position and may probe a range of SF regions of varying ages.
In addition, our best result corresponds to a gaussian kernel which cannot be directly compared to the
exponentials used by \citet{mur06b} in more face-on systems.
Nevertheless, we can consider some representative timescales to see whether 
our kernel size could reasonably represent a link between the CR electron and dust heating scales.

For example, based on spatially resolved images at $\lambda\,20$ cm and $\lambda\,6$ cm, the minimum energy
 magnetic field strength in NGC~3044 has a mean value of
$B\,=\,5.8~\mu$G, ranging from 3.5 to 8.9 $\mu$G  to within uncertainties of approximately
a factor of 2 \citep{irw99}. 
The lifetime for cosmic ray electrons against synchrotron and inverse Compton
losses, for an isotropic velocity distribution of CR electrons,
is given by \citep[e.g.][]{mur06b},
\begin{equation}
\left(\frac{\tau_{CRE}}{\rm yr}\right)\sim \frac{5.7\,\times\,10^7\,
\left(\frac{B}{\mu\rm G}\right)^{0.5}}{
\left(\frac{\nu}{\rm GHz}\right)^{0.5}\left(\frac{u_{B}\,+\,u_{\rm rad}}{10^{-12}\,{\rm ergs~cm^{-3}}}\right)}
\end{equation}
where $B$ is the magnetic field strength and $u_B$ and $u_{\rm rad}$ are the 
magnetic field and radiation energy
densities, respectively.

Taking $u_B\,=\,B^2/(8\pi)$ and  $u_B\,\approx\,u_{rad}$, 
then $\tau\,\propto\,B^{-3/2}$ and we find $\tau\,=\,6.5\,\times\,10^7$ yr
(ranging from to 3.3 to 14$\,\times\,10^7$ yr).  
This implies that the 617 MHz
emission is measuring a massive SF history of NGC~3044 that is integrated over $10^{7-8}$ yr;
 it is not measuring the recent SF activity in the galaxy.

In this time, how far can CR electrons diffuse?  The diffusion timescale can be written \citep{mur09},
\begin{equation}
\left(\frac{\tau_{diff}}{\rm yr}\right)\,=\,2\,\times\,10^6\,\left(\frac{l_{diff}}{{\rm kpc}}\right)^2\,
\left(\frac{\nu}{\rm GHz}\right)^{-0.25}\,\left(\frac{B}{\mu{\rm G}}\right)^{0.25}
\end{equation}
where $l_{diff}$ represents the distance that a CR electron can diffuse from its point of origin via 
a random walk, assuming an energy-dependent diffusion coefficient.  Setting $\tau_{CRE}\,=\,\tau_{diff}$
yields $l_{diff}\,=\,$ 4.3 kpc.  
This is a significant distance and sufficient to allow CR electrons to `fill' the galaxy disk, assuming
that SF has been widely distributed over the past $10^{7-8}$ yr. 

If the cold dust is heated by the ISRF which is similarly widely distributed throughout the disk then,
at least to order of magnitude,
our smoothing kernel should represent roughly the difference between  the CR electron
diffusion length and the dust mean free path to ISRF photons.  The latter is not known for NGC~3044
(and likely varies), but
in our own Galaxy, a typical dust cross-section to 
absorption and scattering in the visible through UV part of the spectrum for big grains,
$\sigma_H$, ranges from 0.2 to 1 $\times\,10^{-21}$ cm$^2$ per hydrogen atom \citep{des90}. 
For this range of dust cross-section and using the mean ISM gas density in NGC~3044 of
$n_H\,\approx\,0.25$ cm$^{-3}$ \citep{lee97}, the 
 mean free path of ISRF photons to
dust is $l_{dust}\,=\,1.3\,\to\,6.5$ kpc.


Our Gaussian smoothing kernel of $\sigma_G\,=\,0.84$ kpc (Table~\ref{tab:fir_rad}) corresponds to
a radius (half-width at half maximum) of $l_G\,=\,$1.0 kpc.  If we convolve this size scale with
$l_{dust}$, the result is $l_{dust-conv}\,=\,1.6\,\to\,6.6$ kpc which can now be compared to 
$l_{diff}\,=\,4.3$ kpc estimated above.
Although there are many uncertainties and assumptions in this comparison,
the result indicates that the interpretation of the smoothing kernel as accounting for the
difference between the CR electron diffusion scale and the photon mean free path to dust is reasonable.

Our results suggest that there is a clear link between synchrotron emission, which measures the integrated massive
SF history in the galaxy over a 
few 10$^{7-8}$ yr, and the ISRF-heated cold dust. 
In the latter case, the fraction of heating that is due to FUV photons from young massive stars is
uncertain \citep[e.g. see the discussion in][]{bel03}.  Older stars contribute approximately an order of
magnitude more photons than OB stars in the Solar neighbourhood \citep[][p. 13]{tie06} whereas the cross-section
for large grains increases by only a factor of 5 from the infrared through the ultraviolet \citep{des90}.
In addition, we are considering emission from regions (of order 4.3 kpc, see above) which are
much larger than typical regions within which hot massive stars exist; the heating source for
the $\lambda\,850~\mu$m dust emission over such size scales is then
the cooler stellar population (as has been found by other authors previously), rather than
the hot, young massive stars that eventually give rise to the synchrotron
emission via supernovae.

What then is the connection and the reason for the non-linear relation?
We first note that non-linear relations have been observed before, 
especially when considering
those of lower luminosity ($L_{IR}\,\ltabouteq\,10^{10}~L_\odot$) and/or galaxies that have been studied at
low radio frequencies ($\nu\,\ltabouteq\,5$ GHz).  In such cases, the relation $L_{radio}\,\propto\,L_{IR}^\gamma$, where
$\gamma\,>\,1$ has been observed.
  Such a trend has been interpreted as
reflecting an increasing contribution from older stars in heating the dust, a non-linear relation
between the SFR and radio continuum emission, or an underproduction of 
both radio and and infrared emission in low luminosity galaxies
\citep[see e.g.][]{bel03}.

 To our knowledge, however, a slope as steep as $\gamma\,=\,2.1$ has
not previously been observed, likely because of the lack of smoothing as described above.
The closest comparisons are to the results of \citet{bra03} and \citet{hoe98}
for NGC~5775 and M~31, respectively, who both find approximately linear relations between non-thermal
radio emission and cold dust, in agreement with our result when a smoothing kernel
is ignored.

For both the 617 MHz and $\lambda\,850~\mu$m emission, we are observing optically thin emission over
equivalent lines of sight.  Then the relation compares the emissivities of the two components
\citep[see also][]{gro03}.  For
synchrotron emission, the emission coefficient can be expressed as,
\begin{equation}
j_{synch}\,\propto\,n_{CRE}\,B^{\frac{\Gamma+1}{2}}
\end{equation}
where $n_{CRE}$ represents number density of CR electrons, $B$ is the magnetic field strength
(we take $B_\perp\,\propto\,B$),
and $\Gamma$ is the power law slope of the CR electron spectrum.

There is substantial observational and theoretical evidence for an approximately constant Alfv{\'e}n speed
in the ISM leading to $B\,\propto\,\sqrt{n_g}$, where $n_g$ is the gas density 
\citep[][and others]{tro86,cha90,gro03,tho09}.  In addition, we expect that
$n_{CRE}\,\propto\,SFR_{m}$, where $SFR_m$ is the massive SFR.  If we accept that the massive SFR is governed
by a Schmidt law such that $SFR_m\,\propto\,n_g^\eta$, then the above equation becomes,
\begin{equation}\label{eqn:synch}
j_{synch}\,\propto\,n_g^{\frac{4\eta\,+\,\Gamma+1}{4}}
\end{equation}

For the dust emissivity, we have,
\begin{equation}
j_{cold~dust}\,\propto\,n_d\,B_\nu(T_c)
\end{equation}
which is essentially a variant of Eqn.~\ref{eqn:modified} in which the optical depth is expressed
explicitly as a function of the number density of dust grains ($n_d$).  If we take a constant gas/dust
ratio, then $n_d\,\propto\,n_g$.  In addition, variations in gas density in the ISM can be many orders
of magnitude, whereas a reasonable variation in the cold dust temperature results in a variation in
$B_\nu(T_c)$ of factors of a few (e.g. increasing the error bar on $T_c$ for NGC~3044 by
a factor of two varies $B_\nu(T_c)$ by a factor of approximately $\pm$ 2). Dust temperatures may vary a great deal
in the ISM, but since we are considering only the $\lambda\,850~\mu$m emission which we have linked to
the cold dust of NGC~3044 only, we will assume that gas density variations dominate and
approximate the cold dust emission as,
 \begin{equation}\label{eqn:cold_dust}
j_{cold~dust}\,\propto\,n_g
\end{equation}
Another way of expressing the above is that the ISRF imposes an essentially constant heating effect
on $j_{cold~dust}$ due to a presumed dominance of widespread and long-lived cooler stars to this heating.

Combining Eqns.~\ref{eqn:synch} and \ref{eqn:cold_dust} and restoring the flux nomenclature yields,
\begin{equation}
S_{617}\,\propto\,S_{850}^{\frac{4\eta\,+\,\Gamma+1}{4}}
\end{equation}
Taking the observational results for the Schmidt law summarized by \cite{ken08}, 
i.e. $\eta\,=\,1.4\,\pm\,0.1$, and using the
mean spectral index between $\lambda\,20$ and $6$ cm  measured by \citet{irw99} for NGC~3044, 
$\bar{\alpha}\,=\,-0.6$, leading to 
$\Gamma\,=\,1\,-\,2\alpha\,=\,2.2$, we find,
\begin{equation}
S_{617}\,\propto\,S_{850}^{2.2}
\end{equation}
which agrees with the measured $S_{617}\,\propto\,S_{850}^{2.1\,\pm\,0.2}$ for NGC~3044, within errors.
 See \cite{nik97} for an earlier example of this kind of approach.

It is important to note the variations that can occur in the above indices.  The measured spectral index of
NGC~3044, for example, ranges from -1.0 to 0.2 within the galaxy and may differ from these values at the
lower frequency that has been used in these observations.  As for the Schmidt law, theoretical considerations
result in values that can generally be found in the range, $\eta\,=\,1\,\to\,2$ \citep{lar92}, though a 
free-fall value yields $\eta\,=\,1.5$ \citep[][his Eqn. 38]{hen91}.  Observationally, variations in $\eta$ are
also observed \citep{big08,ken12}. 
Nevertheless, the above result shows that a simple application of the average spectral index and the most
universally accepted Schmidt law provides a reasonable explanation for the  617 MHz - 850 $\mu$m correlation
found in NGC~3044.  For this result, it is important that
the non-thermal radio emission and cold dust emission are sufficiently isolated spectrally and that the dust scale
length is adjusted through smoothing.

To summarize, if we assume a constant ISRF heating of the cold dust, then the link between synchrotron emission
and cold dust is via the gas density.  A higher gas density yields a higher SFR and a higher
magnetic field, both of which affect the synchrotron emission, and a higher gas density yields a higher dust density
leading to a higher cold dust emission.


\section{Conclusions}
\label{sec:conclusions}

We have observed the edge-on galaxies, NGC~3044 and NGC~4157 at 617 MHz using the GMRT and at $\lambda\,450$ and
$\lambda\,850$ $\mu$m using the JCMT.  These are the first results for these two galaxies at these wavelengths.
The main results are as follows.

$\bullet$  For NGC~3044, high latitude emission is observed at 617 MHz, consistent with previous radio continuum results
and some evidence is present at low intensities for vertical extensions at $\lambda\,850$ $\mu$m also.

$\bullet$ At the far ends of the major axis of NGC~3044, there appear to be disturbances which have resulted in
617 MHz emission extending away from the plane towards the north. The most most obvious of these begin on
the far south-east end of the major axis, including some 'disconnected' features, one of which we have named
the non-thermal cloud (NTC).  
Order-of-magnitude calculations suggest that the NTC could have originated from activity related to star-formation
in the galaxy's disk. 
Another radio continuum extension is seen on the NW end of the major axis at which an HI supershell is known.
Such disturbances will be
more effective at ejecting material away from the plane if they are at
 large galactocentric radii where the disk density is lower.

$\bullet$ Some high latitude 617 MHz emission is observed in NGC~4157 but we do not have the sensitivity to unambiguously
detect halo dust emission at either $\lambda\,450$ $\mu$m or $\lambda\,850$ $\mu$m for this galaxy.  A large 
radio continuum extension
is seen towards the south beginning at the end of the north-eastern major axis in this galaxy and appears to correlate with
high-latitude emission seen at $\lambda\,22$ $\mu$m from the WISE satellite.  

$\bullet$ We could not fit a single temperature model to the sub-mm spectrum for either NGC~3044 or NGC~4157, but could fit
both spectra with a two-temperature model.  For NGC~3044, we find T$_w\,=\,31.0\,\pm\,1.4$ K and
 T$_c\,=\,9.5\,\pm\,1.5$ K and for NGC~4157, we find
T$_w\,=\,25.6\,\pm\,1.4$ K and
 T$_c\,=\,15.3\,\pm\,1.5$ K.  For the latter galaxy, we do not detect emission at the largest galactocentric radii.
Dust masses are $M_d\,=\,(1.6\,\pm\,0.6)\,\times\,10^8~M_\odot$ and 
 $M_d\,=\,(0.21\,\pm\,0.06)\,\times\,10^8~M_\odot$ for NGC~3044 and NGC~4157, respectively.  There is more cold than warm dust
in both galaxies.

$\bullet$  We find a clear correlation between the 617 MHz and   $\lambda\,850$ $\mu$m emission for both NGC~3044 and
NGC~4157.  This correlation improves significantly if a smoothing kernel is applied to the $\lambda\,850$ $\mu$m
data to account for differences between the mean free path of a dust-heating photon and the diffusion length of CR
electrons in the ISM. Simple timescale and length scale arguments suggest that such a smoothing kernel can indeed
account for these differences.

$\bullet$  With an applied smoothing kernel, the 617 MHz and   $\lambda\,850$ $\mu$m correlation becomes 
strongly non-linear.  For the best data set and the one in which the  $\lambda\,850$ $\mu$m is strongly
dominated by cold dust
(i.e. NGC~3044), we find
$S_{617}\,\propto\,S_{850}^{2.1\,\pm\,0.2}$.  

$\bullet$ The non-linear relation between synchrotron emission and cold dust can be understood if the heating
of the cold dust is the ISRF in which cooler stars (rather than hot young stars) dominate, leading to variations
in cold dust emission that are dominated by density rather than temperature variations.  Synchrotron emission
depends on the magnetic field strength and CR electron generation, both of which depend on gas density via
$B\,\propto\,\sqrt{\rho}$ and the Schmidt law, respectively.  With these assumptions, 
$S_{617}\,\propto\,S_{850}^{2.2}$ which agrees with the observed correlation.

\section*{Acknowledgments}

This research has made use of the NASA/IPAC Extragalactic Database (NED) which is operated by the Jet Propulsion Laboratory, California Institute of Technology, under contract with the National Aeronautics and Space Administration.
We are grateful to Dr. Rob Swaters for providing WHISP data and to
Dr. Siow-Wang Lee for providing HI and CO(J=2-1) data.  
Thanks also to R. Rand for providing the H$\alpha$ image.  The CO(J=1-0) data were
originally provided by G. Golla for comparative purposes.  We are especially grateful to Loretta Dunne 
for
insightful comments.


\label{lastpage}

\end{document}